\documentclass[%
 reprint,
 amsmath,amssymb,
 aps,
]{revtex4-2}

\usepackage{graphicx}
\usepackage{dcolumn}
\usepackage{bm}

\begin{document}

\preprint{APS/123-QED}

\title{Cross-stream migration of a vesicle in vortical flows}

\author{G\"{o}kberk Kabacao\u{g}lu}
\affiliation{%
 Department of Mechanical Engineering, \.{I}hsan Do\u{g}ramac{\i} Bilkent University\\
 Ankara, 06800, Turkey}%

\author{Enkeleida Lushi}
\affiliation{
  Department of Mathematical Sciences, New Jersey Institute of Technology\\
 Newark, 07102, NJ, USA}%

\date{\today}

\begin{abstract}
We use numerical simulations to systematically investigate the vesicle dynamics in two-dimensional (2D) Taylor-Green vortex flow in the absence of inertial forces. Vesicles are highly deformable membranes encapsulating an incompressible fluid and they serve as numerical and experimental proxies for biological cells such as red blood cells. Vesicle dynamics has been studied in free-space/bounded shear, Poiseuille and Taylor-Couette flows in 2D and 3D. Taylor-Green vortex are characterized with even more complicated properties than those flows such as non-uniform flow line curvature, shear gradient. We study the effects of three parameters on the vesicle dynamics: the ratio of the interior fluid viscosity to that of the exterior one and the ratio of the shear forces on the vesicle to the membrane stiffness (characterized by the capillary number). Vesicle deformability nonlinearly depends on these parameters. Although the study is in 2D, our findings contribute to the wide spectrum of intriguing vesicle dynamics: vesicles migrate inwards and eventually rotate at the vortex center if they are sufficiently deformable. If not so, they migrate away from the vortex center and travel across the periodic arrays of vortices.
\end{abstract}

\maketitle


\section{\label{sec1}Introduction}

Phospholipid molecules containing a hydrophilic head and a hydrophobic tail come together and form a lipid bilayer which consists in biological membranes~\cite{lipowsky-sackmann95}. The lipid bilayer is fluid but impermeable to many molecules except water molecules. Vesicles are the bilayer sacs that show rich dynamics in their flows even at small velocity and length scales~\cite{lipowsky91,lipowsky-sackmann95}. Vesicles of approximately $10\mu m$ in diameter are numerical and experimental proxies for biological membranes encapsulating only a liquid such as red blood cells. There have been extensive studies on vesicles which are theoretical~\cite{keller-skalak82, seifert99, misbah06, kessler-seifert-e09}, experimental~\cite{fischer-schonbein-e78, abkarian-viallat07, kantsler-steinberg-e08, tomaiuolo-guido-e09, coupier-misbah-e12, dupire-viallat-e12, abkarian-stone-e08, fischer-korzeniewski13, minetti-coupier-e19}, and computational~\cite{queguiner-biesel97, kaoui-misbah09, biben-misbah-e11, zhao-shaqfeh11a, fedosov-gompper-e14, farutin-misbah12, trozzo-jaeger-e15}. They have immense application areas in micro/nano-scale biotechnology: they are used as containers for biochemical reactions~\cite{noireaux-libchaber04,karlsson-orwar-e04} and molecular transport~\cite{fendler80, gregoriadis95}, as vectors for targeted drug delivery~\cite{allen-cullis04, thomas-pandit-e18}.  Vesicles show a wide variety of equilibrium shapes and complex non-equilibrium dynamics in their creeping flows (i.e., the flows where the viscous forces dominate the inertial forces). For a wide range of parameter values vesicles have a symmetric shape called parachute~\cite{skalak-branemark69,chan-liu12} and an asymmetric shape called slipper~\cite{tomaiuolo-guido-e09, kaoui-misbah09, chan-liu12, kihm-quint-e18}. Complicated vesicle dynamics arise from the nonlinear interaction of membrane deformation and fluid flow. Understanding rich vesicle dynamics depending on their deformability helps designing microfluidics devices and techniques for medical diagnoses of diseases~\cite{pivkin-suresh-e16, reichel-fedosov-e19, kabacaoglu-biros19, zhu-brandt-e14}. 

The dynamics of a single vesicle has been studied in several fundamental setups so far: free-space/confined shear~\cite{misbah06,dupire-viallat-e12,gera-spagnolie-e22,schonbein-wells69, goldsmith-macintosh-e72, zhao-shaqfeh09,noguchi-gompper07,kaoui-misbah-e09b} and Poiseuille flows~\cite{abkarian-viallat07,dupire-viallat-e12,biben-misbah-e11,danker-misbah-e09, deschamps-steinberg-e09, kaoui-harting-e12, quint-wagner-e17,guckenberger-gekle-e18,farutin-misbah-e14,farutin-jezewska-e16, kaoui-misbah09,ebrahimi-bagchi-e21,cordasco-bagchi-e14}, Taylor-Couette flow and confined Couette flow~\cite{ghigliotti-misbah-e10}. In those flows, vesicles are observed to show various migration and orientation dynamics stemming from the complicated interplay between the vesicle deformability and the imposed flow characteristics such as the shear rate and the flow line curvature. Due to flow-induced deformation vesicle membrane develops tension so as to keep its arc length constant (due to the inextensibility). Tension along the membrane, then, dictates vesicle dynamics. Vesicle evolves in such a way that minimizes the non-uniformity in tension distribution and at equilibrium tension becomes uniform~\cite{vlahovska-misbah-e09, ghigliotti-misbah-e10}. There are some previously observed dynamics relevant to the present study: (i) membrane tank-treading vs. vesicle tumbling and (ii) cross-stream migration.

In free-space shear flow vesicles have been observed to (1) tank-tread with a stationary angle between its main axis and the flow direction (orientation angle), and (2) tumble, i.e., go through a periodic flipping motion. Which dynamics vesicle shows depends on vesicle's deformability and the ratio of its interior fluid's viscosity to the exterior fluid's viscosity (viscosity contrast) (see~\cite{kantsler-steinberg-e05, kantsler-steinberg-e06, deschamps-steinberg-e09} for the experimental and numerical studies). For low viscosity contrast values, vesicle's membrane tank-treads. The tank-treading motion induces an inner circulation which results in higher dissipation. However, it is shown in 2D free-space Poiseuille flow that tank-treading is a favorable dynamics under specific conditions as it helps vesicle reduces the lag between the vesicle velocity and the imposed flow~\cite{kaoui-misbah09}. As viscosity contrast increases, tank-treading leads to more dissipation and eventually the vesicle transitions to tumbling to reduce the dissipation~\cite{kantsler-steinberg-e05}. The appearance of tank-treading vs. tumbling dynamics is similar in bounded/free Taylor-Couette flow, Poiseuille flow.

Cross-stream migration in low Reynolds number flows may occur if the symmetry in the suspended particle is lost by deformation or in the presence of the wall~\cite{vlahovska-misbah-e09}. Particle deformation is due to a shear gradient (as in free Poiseuille flow~\cite{kaoui-misbah09}) and/or flow line curvature (as in Taylor-Couette flow~\cite{ghigliotti-misbah-e10}). No matter what the source for the deformation is, vesicles are observed to migrate for low viscosity contrast values for which they also tank-tread~\cite{kaoui-misbah09, ghigliotti-misbah-e10}. Above a critical viscosity contrast value, vesicle starts tumbling and the migration is suppressed in free Poiseuille and Taylor-Couette flows. Unlike those free-space flows, in bounded shear flow vesicles migrate even for high but moderate viscosity contrast values~\cite{ouhra-misbah-e18}. In such cases, vesicle initialized near a wall lifts off and also tumbles. The tumbling leads the lift-off angle to decrease and reverse its direction. That leads vesicle to experience pushing force towards the wall. However, since the tumbling cycle results in asymmetric shapes during the two-halves of the tumbling period, vesicle experiences net migration towards the closest wall. For very high viscosity contrast values, vesicle does not even lift off the wall and aligns with the flow near the wall. 

 \begin{figure}
\centering
\includegraphics[scale=0.7]{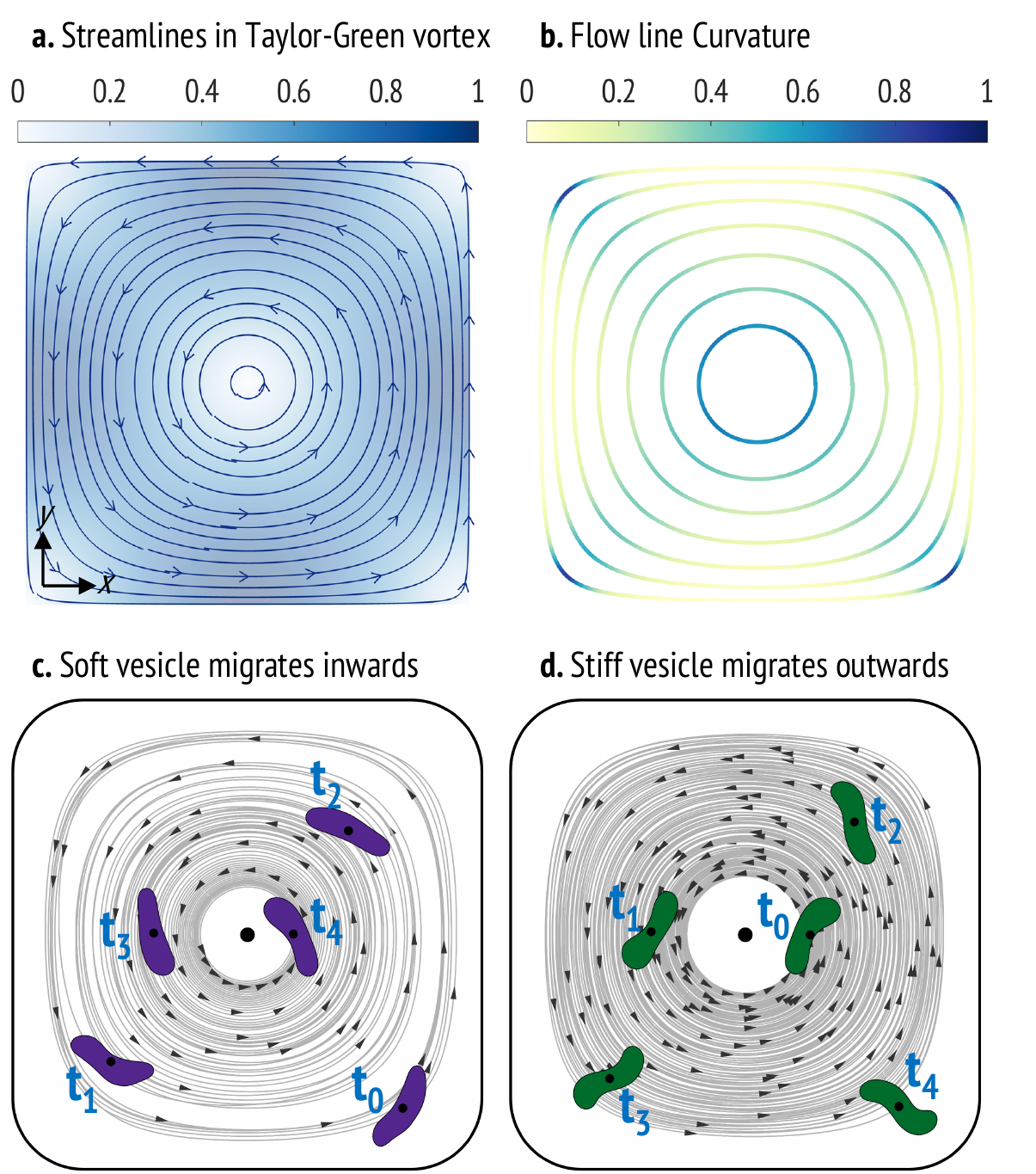}
\caption{\label{fig:fig1} (a) The streamlines with the color scheme showing the background flow speed (nondimensionalized by the imposed flow strength) (b) the flow line curvature in TG flow. (c) Inward migrating vesicle with no viscosity contrast $\lambda = 1$. (d) Outward migrating vesicle with high viscosity contrast value $\lambda = 10$. Both cases have the same capillary number $Ca = 17$ (the flow strength is $400U$). The gray lines show the trajectories of the vesicles. The time of a snapshot is indicated as $t_0$ to $t_4$ . In (c) the vesicle is initialized near the edge of the cell whereas in (d) it is initialized near the vortex center. On one hand, the softer vesicle ($\lambda = 1$) aligns its main axis with the flow lines and migrates inwards. On the other hand, the stiffer one ($\lambda = 10$) tumbles periodically while migrating outwards.}
\end{figure}

Another interesting setup in which the vesicle dynamics needs to be investigated is Taylor-Green (TG) vortex flow~\cite{taylor1923} (see Fig.~\ref{fig:fig1}). It consists of an array of vortices and can be considered a toy model of turbulent flows although it cannot reproduce all the features in turbulent flows and also has some features that are not present in turbulent flows such as closed streamlines. TG flow shows characteristics similar to the flows mentioned above in which vesicle dynamics has been investigated. The flow near the vortex center resembles the Taylor-Couette flow in terms of the flow line curvature and the shear rate. The flow lines in TG flow have almost constant curvature near the vortex center. Further away from the vortex center, the streamlines have non-zero curvature only around the $x=y$ line (see Fig.~\ref{fig:fig1}b). The tangential component of a Taylor-Couette flow is $v_{\theta} \propto 1/r$ and the radial component is zero, which results in curved flow lines whose curvature increases as $r \rightarrow 0$ where is also the high shear rate region. The shear rate reaches its maximum value near the vortex center and disappears near the edges of the the periodic unit of the vortex. 

In this article, we study the transport of a vesicle (a model biological cell) in Taylor-Green vortex flow in a 2D setup in the limit of zero Reynolds number (i.e., the inertial forces are negligible). The vesicle is modeled as inextensible and deformable drop with Helfrich elasticity and its flow is governed by the Stokes equations. We aim at investigating the effects of vesicle deformability on the vesicle dynamics. Specifically, we vary two nondimensional parameters: the capillary number (the ratio of the flow scale to the vesicle's relaxation time scale) and the viscosity contrast (the ratio of the interior fluid viscosity to that of the exterior one). We observe that \textit{sufficiently deformable vesicle migrates towards the vortex center (Fig.~\ref{fig:fig1}c) while it tank-treads and almost aligns its main axis with the imposed flow lines. Whereas stiffer vesicle migrates outwards and travels across the periodic arrays of vortices (Fig.~\ref{fig:fig1}d).} We conducted our studies in two dimensions, motivated by the fact that in several circumstances 2D studies~\cite{kaoui-zimmermann-08} accurately captures the 3D results~\cite{danker-misbah-e09} while 2D simulations are faster and allow mapping parameter space. 

\begin{figure*}
 \centering
 \includegraphics[scale=0.75]{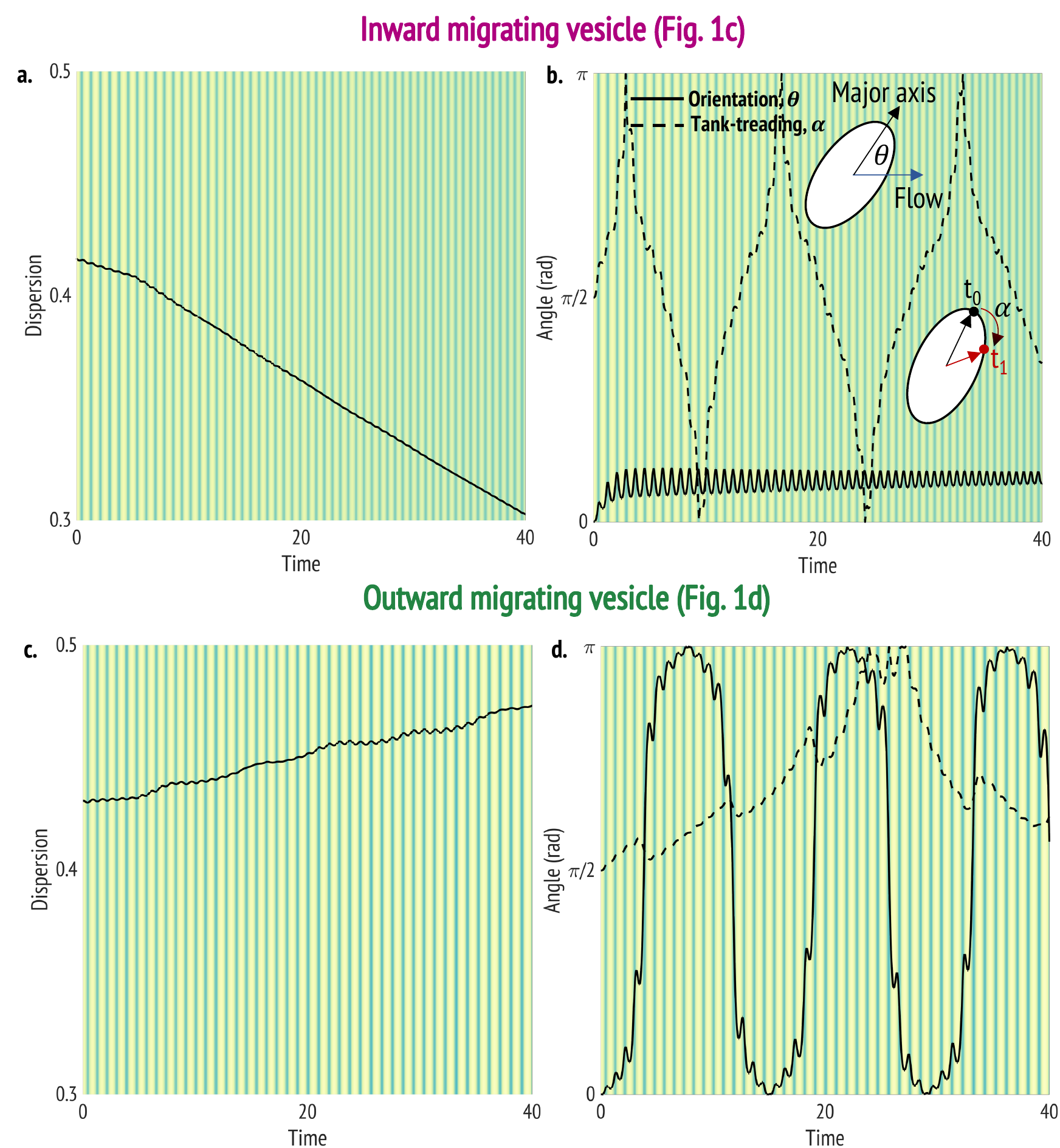}
 \caption{\label{fig:fig3}Dynamics of a vesicle in Taylor-Green flow for low viscosity contrast value ($\lambda = 1$, the first row) and for high viscosity contrast value ($\lambda = 10$, the second row). On the left dispersion, i.e., $L^2$-norm of the distance vector from the vesicle's center to the vortex center is shown. The orientation angle which is defined as the angle between the vesicle's main axis and the imposed velocity vector at the vesicle's center (solid line), and the phase angle of a point on the membrane (i.e., the tank-treading angle) (dashed line) are shown together on the same figures on the right. The background colors in the plots correspond to the the curvature of the flow line where the vesicles are. The color scale for the curvature is given in Fig.~\ref{fig:fig1}b (the yellow is zero curvature and the blue is the maximum curvature). On the one hand (the first row), the inward migrating vesicle has nearly constant orientation angle and its membrane is tank-treading. On the other hand (the second row), the outward migrating vesicle is tumbling with a very slowly tank-treading membrane.}
 \end{figure*}

\section{Methods}\label{sec11}

We consider a vesicle in Taylor-Green vortex flow. The imposed flow is two-dimensional, time-independent and periodic 
\begin{equation}\label{eq:TGflow}
    \mathbf{v}^{\infty}(\mathbf{x}) = U(\sin(x)\cos(y), -\cos(x)\sin(y))
\end{equation}
for $(x,y) \in [0, \pi]^2$ with $U$ the flow strength. The repeated unit flow contains a vortex at the center (Fig.~\ref{fig:fig1}a). A 2x2 array of the units have a hyperbolic stagnation point at its center. That point is connected to other stagnation points through stagnation streamlines (separatrix). We carry out the numerical simulations using a boundary integral formulation (see the Taylor-Couette simulation below and Appendix for the verification and validation of the numerical method used in this study):
\begin{eqnarray}\label{eq:BIEform}
  \mathbf{v}(\mathbf{x}) & = & \frac{2}{1+\lambda}\mathbf{v}^{\infty}(\mathbf{x})\nonumber \\
                          & & + \frac{1}{2\pi\eta_0(1+\lambda)}\int_{\gamma} G(\mathbf{x}-\mathbf{y})\cdot \mathbf{f}(\mathbf{y})\,ds(\mathbf{y}) \nonumber \\
                          & & + \frac{2(1-\lambda)}{\pi(1+\lambda)}\int_{\gamma}\mathbf{v}(\mathbf{y})\cdot T(\mathbf{x}-\mathbf{y})\cdot \mathbf{n}(\mathbf{y}) \, ds(\mathbf{y}),
\end{eqnarray}
where $\gamma$ is the vesicle membrane, $\mathbf{v}$ is the membrane velocity, $G$ and $T$ are the Green's functions of the Stokes flow \cite{kabacaoglu-biros-e18}, $\mathbf{x}$ and $\mathbf{y}$ are points on the membrane, $\mathbf{f}$ is the membrane force/length, $\mathbf{n}$ is the outward normal to the membrane, $\eta_0$ and $\eta_1$ denote the viscosity of the suspending fluid and the fluid inside the vesicle, respectively, and $\lambda = \eta_1/\eta_0$ is the viscosity contrast between the internal and the external fluids. The membrane applies force due to its resistance to bending and its inextensibility. The form of the force/length is obtained by taking the functional derivative of the Helfrich bending energy $E = \frac{\kappa}{2}\int c^2 ds + \int \xi ds$ that includes the tension $\xi$ to enforce the membrane inextensibility:
\begin{equation}\label{eq:membForce}
\mathbf{f}(\mathbf{x}) = -\kappa \left[\frac{d^2 c}{ds^2} + \frac{1}{2}c^3\right]\mathbf{n} + \xi c \mathbf{n} + \frac{d\xi}{ds} \mathbf{t},    
\end{equation}
where $\kappa$ is the membrane's bending modulus, $c$ is the membrane curvature, $\xi$ is the tension that acts like a local Lagrange multiplier enforcing membrane inextensibility, and $\mathbf{t}$ is the tangent to the membrane. The membrane force Eq.~\ref{eq:membForce} balances the jump in the traction across the vesicle membrane. The details of the numerical scheme to solve the integral equation formulation can be found in~\cite{kabacaoglu-biros-e18}.

\begin{figure}
\centering
\includegraphics[scale=0.45]{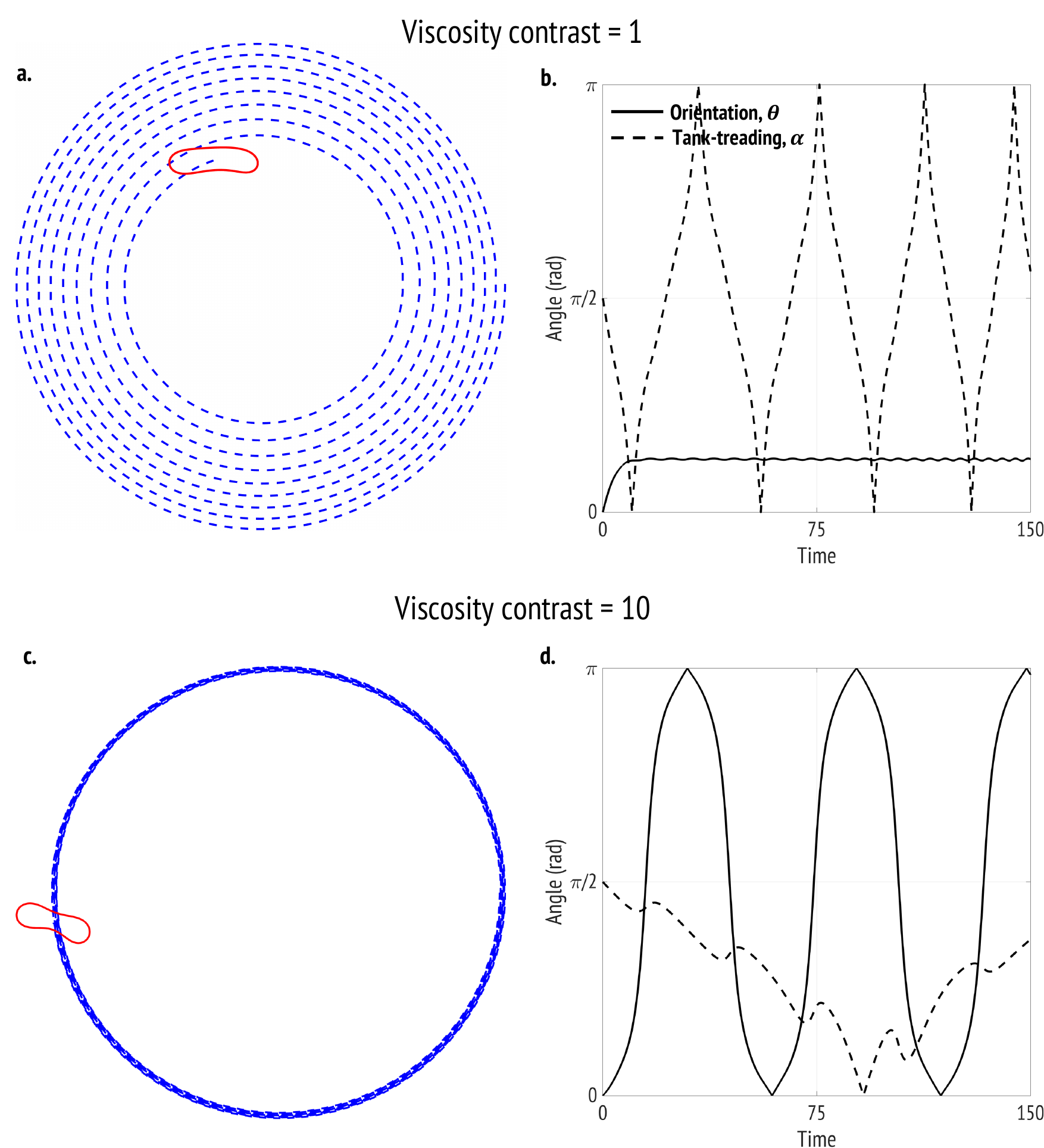}
\caption{\label{fig:fig2} Trajectories of vesicles ($\lambda = 1$ in the first row and $\lambda = 10$ in the second row) in \textbf{Taylor-Couette flow}. The vesicles have similar capillary number and reduced area as those in Fig.~\ref{fig:fig1}. The orientation and the tank-treading angles for those vesicles are also shown on the right. For viscosity contrast $\lambda = 1$, vesicle tank-treads and migrates inwards with a fixed orientation. Whereas for viscosity contrast $\lambda = 10$, it tumbles and shows negligible outward migration.}
\end{figure}

\section{Experiments}\label{sec2}

Let $A$ and $L$ denote the area enclosed by a vesicle and its arclength, respectively. Then, the vesicle's reduced area (deflation) is defined as the ratio of the enclosed area to the area of a circle having the same perimeter $L$:  $\Delta = \sqrt{A/[\pi(L/2\pi)]^2}$ ($0 < \Delta < 1$, $\Delta = 1$ for a circle). The other dimensionless numbers that enter the problem of free-space vesicle flows are (1) the viscosity contrast value $\lambda$, and (2) the capillary number $Ca = \eta_0 U R^3 / \kappa $ where $\kappa$ is the membrane bending stiffness and the vesicle's characteristic size $R$ is defined as the radius of a circle that has the same perimeter as the vesicle, i.e., $R = \sqrt{A/\pi}$. The capillary number measures the vesicle deformability: its higher values correspond to more deformable vesicles.

We experimented numerically by varying the capillary number and the viscosity contrast values. Considering healthy red blood cells, a typical value for velocity in small vessels in human microcirculation is $800 \mu/s$~\cite{fung90}, membrane rigidity is $10^{-19} J$~\cite{mohandas-evans94}, cell radius is $3 \mu m$ and the plasma viscosity is $10^{-3} $ Pa.s. So, typical $Ca$ value is $\mathcal{O}(10)$ for healthy cells. When a cell is diseased, it loses its deformability and $Ca$ becomes one order of magnitude smaller~\cite{kaoui-misbah-e11}. Based on that calculation, we considered $Ca \in [1, 80]$ in our simulations. To observe the dynamics in the limit of rigid vesicle, we reduced $Ca$ to $\mathcal{O}(10^{-2})$, hence worked in the range of $Ca \in [10^{-2}, 80]$. The range of the viscosity contrast values is $\lambda \in [1, 100]$. Reduced area for RBCs is approximately 0.6~\cite{seifert91}. Here, we considered $\Delta = 0.6$ for which the ratio of the vortex size to the vesicle size $R$ is 18. 

To analyze our simulations, we investigate several quantities. One of them is the dispersion of a vesicle which is defined as $L^2$-norm of the distance vector from the vesicle's center to the vortex center. Second one is the orientation angle $\theta$ of a vesicle with respect to the imposed flow at its center. We quantify a vesicle's orientation by the angle between its main axis and the velocity vector at its center (see the inset in Fig.~\ref{fig:fig3}b). To do so, we rotate vesicle such that the imposed velocity at the center is in $x$ direction. Then, the main axis of a vesicle is the axis corresponding to the smallest principal moment of inertial with the $x$-axis. The moment of inertia tensor $J$ is 
\begin{equation*}
    J = \int_{\omega} \, (\|\mathbf{r}\|^2 I - \mathbf{r} \otimes \mathbf{r})\, d\mathbf{x} = \frac{1}{4} \int_{\gamma} \, \mathbf{r}\cdot\mathbf{n}(\|\mathbf{r}\|^2 I - \mathbf{r} \otimes \mathbf{r})\,ds,
\end{equation*}
where $\omega$ is the area enclosed by $\gamma$, $\gamma$ denotes vesicle membrane, and $\mathbf{r} = \mathbf{x}-\mathbf{c}$ is the distance of point $\mathbf{x}$ from the vesicle's center $\mathbf{c}$. The principal axes of inertia are the eigenvectors of $J$. The last quantity is the tank-treading angle $\alpha$ that measures the angular position of a particular point on the membrane with respect to the vesicle's center (see the inset in Fig.~\ref{fig:fig3}). Since a vesicle rotates and translates with respect to the vortex center in TG flow, the tank-treading angle is also measured after aligning the vesicle such that the imposed flow at its center is in the $x$-direction.

 \begin{figure}
 \centering
 \includegraphics[scale=0.8]{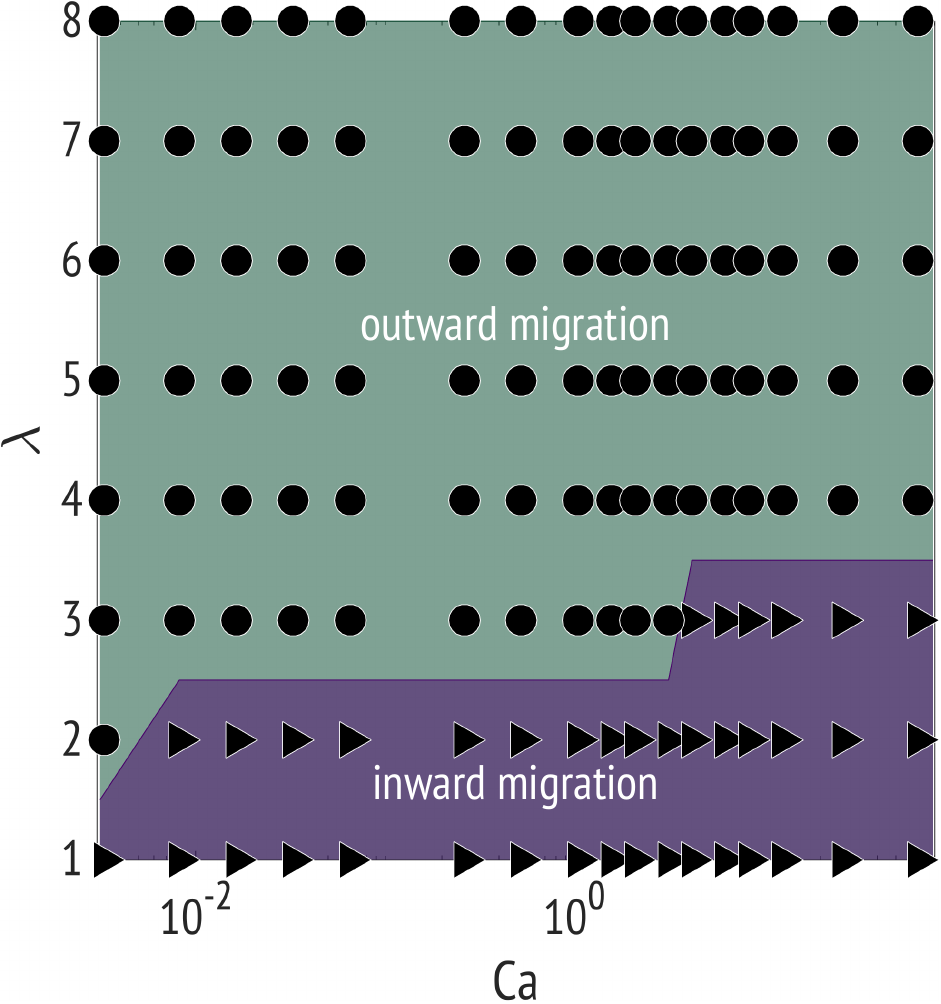}
 \caption{\label{fig:fig4}Phase diagram for a vesicle with reduced area $\Delta = 0.6$. $Ca$ is the capillary number ranging from $\mathcal{O}(10^{-2})$ to almost 80 and $\lambda$ is the viscosity contrast. The purple area shows inward migration and the green area shows outward migration. The filled circles and triangles represent tumbling and tank-treading dynamics, respectively.}
 \end{figure}
 
\section{Discussion}
{\it Tank-treading vesicles migrate inwards.} In Fig.~\ref{fig:fig1}c, we superimpose the snapshots from the simulation of a vesicle with $\lambda = 1$ initialized near the edge of the periodic unit. The vesicle's reduced area is $\Delta = 0.6$ and the capillary number is $Ca = 17$. The vesicle with low viscosity contrast value  migrate inwards (Fig.~\ref{fig:fig3}a) all the while it tank-treads with almost a fixed orientation angle with respect to the imposed flow direction at the vesicle's center (Fig.~\ref{fig:fig3}b). This is similar to the vesicle dynamics in Taylor-Couette flow for low viscosity contrast values~\cite{ghigliotti-misbah-e10}. We also performed simulations by initializing a vesicle closer to the edge of the repeating unit (only 5\% of the vortex size away from the edge) for the same viscosity contrast value. We observed persistent inward migration regardless of the initial position and orientation of the vesicle.

{\it Tumbling vesicles migrate outwards.} Under the same flow conditions, we, then, increased the viscosity contrast of the vesicle to $\lambda = 10$. Figure~\ref{fig:fig1}d shows the snapshots from this simulation. Here, the vesicle is initialized near the vortex center and migrates outwards (see Fig.~\ref{fig:fig3}c for its dispersion). At the same time, the vesicle tumbles - that can be captured by looking at its orientation angle in Fig.~\ref{fig:fig3}d. Its membrane still tank-treads but much slower than it does for low viscosity contrast values. Outward migration with tumbling for high viscosity contrast values is a new phenomenon, which is not observed in Taylor-Couette flows~\cite{ghigliotti-misbah-e10}.

{\it Flow line curvature impacts vesicle dynamics.} To illustrate that, in Fig.~\ref{fig:fig3} we plot the dispersion, the orientation and the tank-treading angles. Time is in the $x$-axes and nondimensionalized with the time scale (the time scale is $\pi/U$ where $\pi$ is the size of a repeating unit and $U$ is the magnitude of the TG flow in Eq.~\ref{eq:TGflow}). The striped background in these plots indicates the curvature of the flow line the vesicle resides at a particular time. The color scale is given in Fig.~\ref{fig:fig1}b (the yellow is zero curvature and the blue is the maximum curvature). For the inward migrating vesicle (the first row in Fig.~\ref{fig:fig3}), its orientation angle drifts away from the flow direction when it is on flow line of a lower curvature (yellow bands in Fig.~\ref{fig:fig3}b). Its membrane tank-treads faster on high curvature flow lines than on low curvature flow lines. Although the flow line curvature does not have a significant impact on the dispersion of the outward migrating vesicle, its effects on the vesicle orientation and tank-treading are visible. High flow line curvature accelerates the vesicle's tumbling and tank-treading. 

Taylor-Couette flow resembles Taylor-Green flow in several aspects such as having non-zero flow line curvature and high shear rate at the center~\cite{ghigliotti-misbah-e10}. A fundamental difference is that flow line curvature in TG flow varies along a streamline. While vesicles migrate inwards for low viscosity contrast values in both flows, for high viscosity contrast values they show negligible migration in Taylor-Couette flow and outward migration in TG flow. In order to understand the reasons for the different dynamics, we considered a single vesicle in Taylor-Couette flow where the imposed flow is $v_{\theta} = a/r$, $v_r = 0$ where $r$ is the position of vesicle's center. Then, the imposed shear rate becomes $-2a/r$. Vesicle has the reduced area $\Delta = 0.6$ (same as in the TG flow simulations in Fig.~\ref{fig:fig1}). We initialized the vesicle at $10R$ and the imposed flow has $a = -40$. So, the capillary number is 0.8 at $10R$ and increases to 80 at $R$. We performed two simulations for different viscosity contrast values $\lambda = (1, 10)$. Our results shown in Fig.~\ref{fig:fig2} recapitulated the findings in~\cite{ghigliotti-misbah-e10}. For $\lambda = 1$, vesicle migrates inwards (see its trajectory on the left in the first row) with a fixed orientation (the solid line on the right in the first row) while tank-treading. The same dynamics is also observed in TG flow. However, for $\lambda = 10$, vesicle does not migrate (see its trajectory on the right in the second row). That vesicle tumbles periodically (see the solid line in the right figure in the second row) with slight tank-treading. Although tumbling observed in both Taylor-Couette and TG flows, the time evolution of the orientation angle is different. While the orientation smoothly changes in Taylor-Couette flow, the rate of change is varying as the flow line curvature varies in TG flow. The non-uniformity in the flow line curvature in TG flow inhibits symmetric tumbling as in Taylor-Couette flow and hence leads to non-negligible outward migration. \textit{Does a tumbling vesicle always migrate outwards in Taylor-Green vortex?} Even when a vesicle is initialized only 5\% of the vortex size away from the vortex center, it still migrates outwards for the same viscosity contrast value. 

{\it Large $Ca$ value leads to inward migration.}  Since vesicle is more deformable for lower viscosity contrast values and we observe that vesicles migrate inwards for low viscosity contrast values, one would expect inward migration for large capillary number values. Our results shown in the phase diagram (Fig.~\ref{fig:fig4}) verify that. As the capillary number increases, the critical viscosity contrast value for the transition from inward migration to outward migration increases initially and then the migration direction does not significantly depend on $Ca$. In free space shear flow, the critical viscosity contrast value for the tank-treading/tumbling transition depends similarly on the capillary number~\cite{kaoui-misbah-e09b}. Figure~\ref{fig:fig4} shows that the tank-treading vs. tumbling dynamics in Taylor-Green flow always coincides with the migration dynamics.


\section{Conclusion}
The present study systematically uncovers complex vesicle dynamics in a complicated Taylor-Green flow (e.g., showing oscillatory flow line curvatures). What is new to the vesicle dynamics in free-space flows is the outward migration of tumbling vesicles for high viscosity contrast values for which vesicles do not significantly migrate in free-space shear and Poiseuille flows. The non-uniformity of flow line curvature along a streamline in TG flow causes symmetry-breaking in the tumbling motion of vesicles, which leads them to migrate away from the vortex center and eventually to travel across the arrays of TG vortices. Taylor-Green vortex is a three-dimensional phenomenon and we expect even more complicated vesicle dynamics in 3D TG flows as similar extensions of shear/Poiseuille flows to 3D have discovered rich dynamics~\cite{biben-misbah-e11,agarwal-biros20}.



\appendix*
\section{Validation of the Numerical Method}
\begin{figure}
\centering
\includegraphics[scale=0.6]{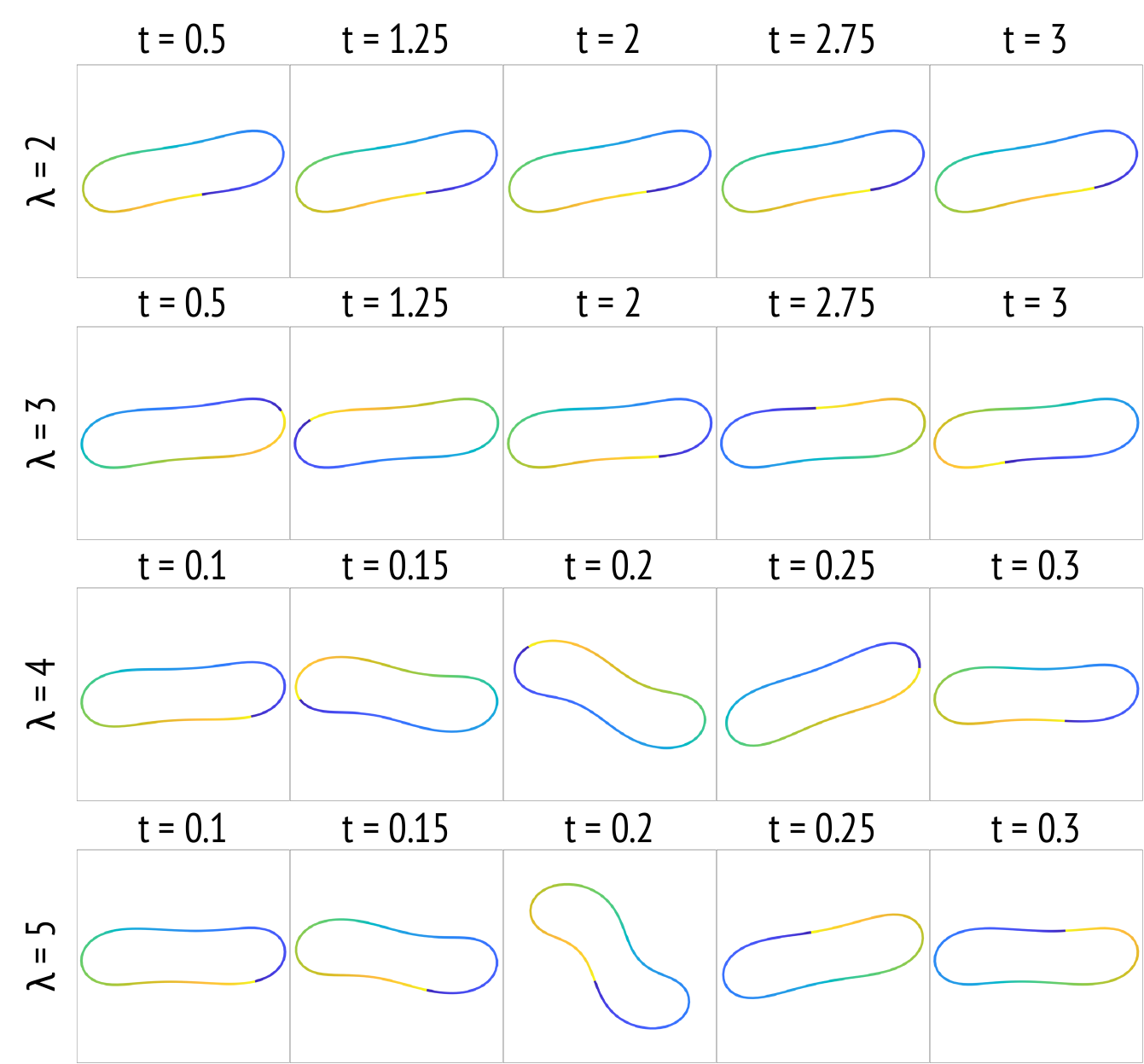}
\caption{\label{fig:fig5} The evolution of a single vesicle in an unbounded shear flow. The viscosity contrast values are $\lambda = 2, 3, 4, $ and $5$. The capillary number is $Ca = 2$. In theory~\cite{kaoui-misbah-e09b} the critical viscosity contrast value for the transition to tumbling is found to be between 3 and 4. Lagrangian points on the vesicle membrane is colored for visual purposes only and has no other significance.}
\end{figure}
We refer the reader to~\cite{kabacaoglu-biros-e18} for the equations governing vesicle flows, their integral equation formulation and the solution techniques used in the boundary integral equation method. In the same article, the method was also verified. In~\cite{kabacaoglu-biros19}, the method was used to accurately reproduce some fundamental results in the literature: capturing the equilibrium shape of a single vesicle in free-space Poiseuille flow with no viscosity contrast, cross-stream migration of vesicles in Taylor-Couette flow. 

In this work, we verified the method by simulating the outward migrating vesicle in Fig.~\ref{fig:fig1}d by refining the spatio-temporal resolution. The time scale (the ratio of the vortex size to the flow strength $U$) in this test problem is 0.02s. We reduced the time step size to $10^{-7}s$ and the outward migration is still observed. To validate the method, we reproduced the cross-stream migration results in Taylor-Couette flow which are presented in the main text. In addition to that, we performed simulations of a single vesicle in free-space shear flow in which vesicle tank-treads at an equilibrium angle for low viscosity contrast values and tumbles otherwise~\cite{kaoui-misbah-e09b, kantsler-steinberg-e06, misbah06}. The critical viscosity contrast value depends on the reduced area. For reduced area of $\Delta = 0.6$, the critical viscosity contrast value is between 3 and 4~\cite{kaoui-misbah-e09b, rahimian-biros-e10}. We considered viscosity contrast values $\lambda$ between 2 and 5 (see Fig.~\ref{fig:fig5}) to capture the transition from tank-treading to tumbling. We set the reduced area to 0.6 and capillary number to 2. We found that vesicle tank-treads for $\lambda < 4$ and tumbles otherwise (see Fig. showing snapshots from the simulations). So, our method accurately captures the critical viscosity contrast value. Additionally, our results show that as the viscosity contrast value increases, vesicle's main axis gets more aligned with the flow direction, which is physically correct.

\bibliography{apssamp}

\providecommand{\noopsort}[1]{}\providecommand{\singleletter}[1]{#1}%
\begin{thebibliography}{64}%
\makeatletter
\providecommand \@ifxundefined [1]{%
 \@ifx{#1\undefined}
}%
\providecommand \@ifnum [1]{%
 \ifnum #1\expandafter \@firstoftwo
 \else \expandafter \@secondoftwo
 \fi
}%
\providecommand \@ifx [1]{%
 \ifx #1\expandafter \@firstoftwo
 \else \expandafter \@secondoftwo
 \fi
}%
\providecommand \natexlab [1]{#1}%
\providecommand \enquote  [1]{``#1''}%
\providecommand \bibnamefont  [1]{#1}%
\providecommand \bibfnamefont [1]{#1}%
\providecommand \citenamefont [1]{#1}%
\providecommand \href@noop [0]{\@secondoftwo}%
\providecommand \href [0]{\begingroup \@sanitize@url \@href}%
\providecommand \@href[1]{\@@startlink{#1}\@@href}%
\providecommand \@@href[1]{\endgroup#1\@@endlink}%
\providecommand \@sanitize@url [0]{\catcode `\\12\catcode `\$12\catcode
  `\&12\catcode `\#12\catcode `\^12\catcode `\_12\catcode `\%12\relax}%
\providecommand \@@startlink[1]{}%
\providecommand \@@endlink[0]{}%
\providecommand \url  [0]{\begingroup\@sanitize@url \@url }%
\providecommand \@url [1]{\endgroup\@href {#1}{\urlprefix }}%
\providecommand \urlprefix  [0]{URL }%
\providecommand \Eprint [0]{\href }%
\providecommand \doibase [0]{https://doi.org/}%
\providecommand \selectlanguage [0]{\@gobble}%
\providecommand \bibinfo  [0]{\@secondoftwo}%
\providecommand \bibfield  [0]{\@secondoftwo}%
\providecommand \translation [1]{[#1]}%
\providecommand \BibitemOpen [0]{}%
\providecommand \bibitemStop [0]{}%
\providecommand \bibitemNoStop [0]{.\EOS\space}%
\providecommand \EOS [0]{\spacefactor3000\relax}%
\providecommand \BibitemShut  [1]{\csname bibitem#1\endcsname}%
\let\auto@bib@innerbib\@empty
\bibitem [{\citenamefont {Lipowsky}\ and\ \citenamefont
  {Sackmann}(1995)}]{lipowsky-sackmann95}%
  \BibitemOpen
  \bibfield  {author} {\bibinfo {author} {\bibfnamefont {R.}~\bibnamefont
  {Lipowsky}}\ and\ \bibinfo {author} {\bibfnamefont {E.}~\bibnamefont
  {Sackmann}},\ }\href@noop {} {\emph {\bibinfo {title} {{Structure and
  Dynamics of Membranes}}}}\ (\bibinfo  {publisher} {Elsevier},\ \bibinfo
  {address} {Amsterdam},\ \bibinfo {year} {1995})\BibitemShut {NoStop}%
\bibitem [{\citenamefont {Lipowsky}(1991)}]{lipowsky91}%
  \BibitemOpen
  \bibfield  {author} {\bibinfo {author} {\bibfnamefont {R.}~\bibnamefont
  {Lipowsky}},\ }\bibfield  {title} {\bibinfo {title} {{The conformation of
  membranes}},\ }\href@noop {} {\bibfield  {journal} {\bibinfo  {journal}
  {Nature}\ }\textbf {\bibinfo {volume} {349}},\ \bibinfo {pages} {475}
  (\bibinfo {year} {1991})}\BibitemShut {NoStop}%
\bibitem [{\citenamefont {Keller}\ and\ \citenamefont
  {Skalak}(1982)}]{keller-skalak82}%
  \BibitemOpen
  \bibfield  {author} {\bibinfo {author} {\bibfnamefont {S.~R.}\ \bibnamefont
  {Keller}}\ and\ \bibinfo {author} {\bibfnamefont {R.}~\bibnamefont
  {Skalak}},\ }\bibfield  {title} {\bibinfo {title} {Motion of a tank-treading
  ellipsoidal particle in a shear flow},\ }\href@noop {} {\bibfield  {journal}
  {\bibinfo  {journal} {Journal of Fluid Mechanics}\ }\textbf {\bibinfo
  {volume} {120}},\ \bibinfo {pages} {27} (\bibinfo {year} {1982})}\BibitemShut
  {NoStop}%
\bibitem [{\citenamefont {Seifert}(1999)}]{seifert99}%
  \BibitemOpen
  \bibfield  {author} {\bibinfo {author} {\bibfnamefont {U.}~\bibnamefont
  {Seifert}},\ }\bibfield  {title} {\bibinfo {title} {Fluid membranes in
  hydrodynamic flow fields: formalism and application to fluctuating
  quasispherical vesicles in shear flow},\ }\href@noop {} {\bibfield  {journal}
  {\bibinfo  {journal} {Eur. Phys. J. B}\ }\textbf {\bibinfo {volume} {8}},\
  \bibinfo {pages} {405} (\bibinfo {year} {1999})}\BibitemShut {NoStop}%
\bibitem [{\citenamefont {Misbah}(2006)}]{misbah06}%
  \BibitemOpen
  \bibfield  {author} {\bibinfo {author} {\bibfnamefont {C.}~\bibnamefont
  {Misbah}},\ }\bibfield  {title} {\bibinfo {title} {{Vacillating breathing and
  tumbling of vesicles under shear flow}},\ }\href@noop {} {\bibfield
  {journal} {\bibinfo  {journal} {Physical Review Letters}\ }\textbf {\bibinfo
  {volume} {96}} (\bibinfo {year} {2006})}\BibitemShut {NoStop}%
\bibitem [{\citenamefont {Kessler}\ \emph {et~al.}(2009)\citenamefont
  {Kessler}, \citenamefont {Finken},\ and\ \citenamefont
  {Seifert}}]{kessler-seifert-e09}%
  \BibitemOpen
  \bibfield  {author} {\bibinfo {author} {\bibfnamefont {S.}~\bibnamefont
  {Kessler}}, \bibinfo {author} {\bibfnamefont {R.}~\bibnamefont {Finken}},\
  and\ \bibinfo {author} {\bibfnamefont {U.}~\bibnamefont {Seifert}},\
  }\bibfield  {title} {\bibinfo {title} {Elastic capsules in shear flow:
  Analytical solutions for constant and time dependent shear rates},\
  }\href@noop {} {\bibfield  {journal} {\bibinfo  {journal} {Eur. Phys. J. E}\
  }\textbf {\bibinfo {volume} {29}},\ \bibinfo {pages} {399} (\bibinfo {year}
  {2009})}\BibitemShut {NoStop}%
\bibitem [{\citenamefont {Fischer}\ \emph {et~al.}(1978)\citenamefont
  {Fischer}, \citenamefont {Stohr-Lissen},\ and\ \citenamefont
  {Schmid-Schonbein}}]{fischer-schonbein-e78}%
  \BibitemOpen
  \bibfield  {author} {\bibinfo {author} {\bibfnamefont {T.~M.}\ \bibnamefont
  {Fischer}}, \bibinfo {author} {\bibfnamefont {M.}~\bibnamefont
  {Stohr-Lissen}},\ and\ \bibinfo {author} {\bibfnamefont {H.}~\bibnamefont
  {Schmid-Schonbein}},\ }\bibfield  {title} {\bibinfo {title} {The red cell as
  a fluid droplet: Tank tread-like motion of the human erythrocyte membrane in
  shear flow},\ }\href@noop {} {\bibfield  {journal} {\bibinfo  {journal}
  {Science}\ }\textbf {\bibinfo {volume} {202}},\ \bibinfo {pages} {894}
  (\bibinfo {year} {1978})}\BibitemShut {NoStop}%
\bibitem [{\citenamefont {Abkarian}\ \emph {et~al.}(2007)\citenamefont
  {Abkarian}, \citenamefont {Faivre},\ and\ \citenamefont
  {Viallat}}]{abkarian-viallat07}%
  \BibitemOpen
  \bibfield  {author} {\bibinfo {author} {\bibfnamefont {M.}~\bibnamefont
  {Abkarian}}, \bibinfo {author} {\bibfnamefont {M.}~\bibnamefont {Faivre}},\
  and\ \bibinfo {author} {\bibfnamefont {A.}~\bibnamefont {Viallat}},\
  }\bibfield  {title} {\bibinfo {title} {Swinging of red blood cells under
  shear flow},\ }\href@noop {} {\bibfield  {journal} {\bibinfo  {journal}
  {Physical Review Letters}\ }\textbf {\bibinfo {volume} {98}},\ \bibinfo
  {pages} {188302} (\bibinfo {year} {2007})}\BibitemShut {NoStop}%
\bibitem [{\citenamefont {Kantsler}\ \emph {et~al.}(2008)\citenamefont
  {Kantsler}, \citenamefont {Segre},\ and\ \citenamefont
  {Steinberg}}]{kantsler-steinberg-e08}%
  \BibitemOpen
  \bibfield  {author} {\bibinfo {author} {\bibfnamefont {V.}~\bibnamefont
  {Kantsler}}, \bibinfo {author} {\bibfnamefont {E.}~\bibnamefont {Segre}},\
  and\ \bibinfo {author} {\bibfnamefont {V.}~\bibnamefont {Steinberg}},\
  }\bibfield  {title} {\bibinfo {title} {Dynamics of interacting vesicles and
  rheology of vesicle suspension in shear flow},\ }\href@noop {} {\bibfield
  {journal} {\bibinfo  {journal} {Europhys. Lett}\ }\textbf {\bibinfo {volume}
  {82}},\ \bibinfo {pages} {58005} (\bibinfo {year} {2008})}\BibitemShut
  {NoStop}%
\bibitem [{\citenamefont {Tomaiuolo}\ \emph {et~al.}(2009)\citenamefont
  {Tomaiuolo}, \citenamefont {Simeone}, \citenamefont {Martinelli},
  \citenamefont {Rotoli},\ and\ \citenamefont {Guido}}]{tomaiuolo-guido-e09}%
  \BibitemOpen
  \bibfield  {author} {\bibinfo {author} {\bibfnamefont {G.}~\bibnamefont
  {Tomaiuolo}}, \bibinfo {author} {\bibfnamefont {M.}~\bibnamefont {Simeone}},
  \bibinfo {author} {\bibfnamefont {V.}~\bibnamefont {Martinelli}}, \bibinfo
  {author} {\bibfnamefont {B.}~\bibnamefont {Rotoli}},\ and\ \bibinfo {author}
  {\bibfnamefont {S.}~\bibnamefont {Guido}},\ }\bibfield  {title} {\bibinfo
  {title} {Red blood cell deformation in microconfined flow},\ }\href@noop {}
  {\bibfield  {journal} {\bibinfo  {journal} {Soft Matter}\ }\textbf {\bibinfo
  {volume} {5}},\ \bibinfo {pages} {3736} (\bibinfo {year} {2009})}\BibitemShut
  {NoStop}%
\bibitem [{\citenamefont {Coupier}\ \emph {et~al.}(2012)\citenamefont
  {Coupier}, \citenamefont {Farutin}, \citenamefont {Minetti}, \citenamefont
  {Podgorski},\ and\ \citenamefont {Misbah}}]{coupier-misbah-e12}%
  \BibitemOpen
  \bibfield  {author} {\bibinfo {author} {\bibfnamefont {G.}~\bibnamefont
  {Coupier}}, \bibinfo {author} {\bibfnamefont {A.}~\bibnamefont {Farutin}},
  \bibinfo {author} {\bibfnamefont {C.}~\bibnamefont {Minetti}}, \bibinfo
  {author} {\bibfnamefont {T.}~\bibnamefont {Podgorski}},\ and\ \bibinfo
  {author} {\bibfnamefont {C.}~\bibnamefont {Misbah}},\ }\bibfield  {title}
  {\bibinfo {title} {Shape diagram of vesicles in poiseuille flow},\
  }\href@noop {} {\bibfield  {journal} {\bibinfo  {journal} {Phys. Rev. Lett.}\
  }\textbf {\bibinfo {volume} {108}},\ \bibinfo {pages} {178106} (\bibinfo
  {year} {2012})}\BibitemShut {NoStop}%
\bibitem [{\citenamefont {Dupire}\ \emph {et~al.}(2012)\citenamefont {Dupire},
  \citenamefont {Socol},\ and\ \citenamefont {Viallat}}]{dupire-viallat-e12}%
  \BibitemOpen
  \bibfield  {author} {\bibinfo {author} {\bibfnamefont {J.}~\bibnamefont
  {Dupire}}, \bibinfo {author} {\bibfnamefont {M.}~\bibnamefont {Socol}},\ and\
  \bibinfo {author} {\bibfnamefont {A.}~\bibnamefont {Viallat}},\ }\bibfield
  {title} {\bibinfo {title} {Full dynamics of red blood cell in shear flow},\
  }\href@noop {} {\bibfield  {journal} {\bibinfo  {journal} {Proc. Natl. Acad.
  Sci. USA}\ }\textbf {\bibinfo {volume} {109}},\ \bibinfo {pages} {20808}
  (\bibinfo {year} {2012})}\BibitemShut {NoStop}%
\bibitem [{\citenamefont {Abkarian}\ \emph {et~al.}(2008)\citenamefont
  {Abkarian}, \citenamefont {Faivre}, \citenamefont {Horton}, \citenamefont
  {Smistrup}, \citenamefont {Best-Popescu},\ and\ \citenamefont
  {Stone}}]{abkarian-stone-e08}%
  \BibitemOpen
  \bibfield  {author} {\bibinfo {author} {\bibfnamefont {M.}~\bibnamefont
  {Abkarian}}, \bibinfo {author} {\bibfnamefont {M.}~\bibnamefont {Faivre}},
  \bibinfo {author} {\bibfnamefont {R.}~\bibnamefont {Horton}}, \bibinfo
  {author} {\bibfnamefont {K.}~\bibnamefont {Smistrup}}, \bibinfo {author}
  {\bibfnamefont {C.~A.}\ \bibnamefont {Best-Popescu}},\ and\ \bibinfo {author}
  {\bibfnamefont {H.~A.}\ \bibnamefont {Stone}},\ }\bibfield  {title} {\bibinfo
  {title} {Cellular-scale hydrodynamics},\ }\href@noop {} {\bibfield  {journal}
  {\bibinfo  {journal} {Biomed. Mater.}\ }\textbf {\bibinfo {volume} {3}},\
  \bibinfo {pages} {034011} (\bibinfo {year} {2008})}\BibitemShut {NoStop}%
\bibitem [{\citenamefont {Fischer}\ and\ \citenamefont
  {Korzeniewski}(2013)}]{fischer-korzeniewski13}%
  \BibitemOpen
  \bibfield  {author} {\bibinfo {author} {\bibfnamefont {T.~M.}\ \bibnamefont
  {Fischer}}\ and\ \bibinfo {author} {\bibfnamefont {R.}~\bibnamefont
  {Korzeniewski}},\ }\bibfield  {title} {\bibinfo {title} {Threshold shear
  stress for the transition between tumbling and tank- treading of red blood
  cells in shear flow: Dependence on the viscosity of the suspending medium},\
  }\href@noop {} {\bibfield  {journal} {\bibinfo  {journal} {Journal of Fluid
  Mechanics}\ }\textbf {\bibinfo {volume} {736}},\ \bibinfo {pages} {351}
  (\bibinfo {year} {2013})}\BibitemShut {NoStop}%
\bibitem [{\citenamefont {Minetti}\ \emph {et~al.}(2019)\citenamefont
  {Minetti}, \citenamefont {Audemar}, \citenamefont {Podgorski},\ and\
  \citenamefont {Coupier}}]{minetti-coupier-e19}%
  \BibitemOpen
  \bibfield  {author} {\bibinfo {author} {\bibfnamefont {C.}~\bibnamefont
  {Minetti}}, \bibinfo {author} {\bibfnamefont {V.}~\bibnamefont {Audemar}},
  \bibinfo {author} {\bibfnamefont {T.}~\bibnamefont {Podgorski}},\ and\
  \bibinfo {author} {\bibfnamefont {G.}~\bibnamefont {Coupier}},\ }\bibfield
  {title} {\bibinfo {title} {Dynamics of a large population of red blood cells
  under shear flow},\ }\href@noop {} {\bibfield  {journal} {\bibinfo  {journal}
  {Journal of Fluid Mechanics}\ }\textbf {\bibinfo {volume} {864}},\ \bibinfo
  {pages} {408} (\bibinfo {year} {2019})}\BibitemShut {NoStop}%
\bibitem [{\citenamefont {Queguiner}\ and\ \citenamefont
  {Barthes-Biesel}(2019)}]{queguiner-biesel97}%
  \BibitemOpen
  \bibfield  {author} {\bibinfo {author} {\bibfnamefont {C.}~\bibnamefont
  {Queguiner}}\ and\ \bibinfo {author} {\bibfnamefont {D.}~\bibnamefont
  {Barthes-Biesel}},\ }\bibfield  {title} {\bibinfo {title} {Axisymmetric
  motion of capsules through cylindrical channels},\ }\href@noop {} {\bibfield
  {journal} {\bibinfo  {journal} {Journal of Fluid Mechanics}\ }\textbf
  {\bibinfo {volume} {864}},\ \bibinfo {pages} {408} (\bibinfo {year}
  {2019})}\BibitemShut {NoStop}%
\bibitem [{\citenamefont {Kaoui}\ \emph
  {et~al.}(2009{\natexlab{a}})\citenamefont {Kaoui}, \citenamefont {Biros},\
  and\ \citenamefont {Misbah}}]{kaoui-misbah09}%
  \BibitemOpen
  \bibfield  {author} {\bibinfo {author} {\bibfnamefont {B.}~\bibnamefont
  {Kaoui}}, \bibinfo {author} {\bibfnamefont {G.}~\bibnamefont {Biros}},\ and\
  \bibinfo {author} {\bibfnamefont {C.}~\bibnamefont {Misbah}},\ }\bibfield
  {title} {\bibinfo {title} {{Why Do Red Blood Cells Have Asymmetric Shapes
  Even in a Symmetric Flow?}},\ }\href@noop {} {\bibfield  {journal} {\bibinfo
  {journal} {Physical Review Letters}\ }\textbf {\bibinfo {volume} {103}},\
  \bibinfo {pages} {188101} (\bibinfo {year} {2009}{\natexlab{a}})}\BibitemShut
  {NoStop}%
\bibitem [{\citenamefont {Biben}\ \emph {et~al.}(2011)\citenamefont {Biben},
  \citenamefont {Farutin},\ and\ \citenamefont {Misbah}}]{biben-misbah-e11}%
  \BibitemOpen
  \bibfield  {author} {\bibinfo {author} {\bibfnamefont {T.}~\bibnamefont
  {Biben}}, \bibinfo {author} {\bibfnamefont {A.}~\bibnamefont {Farutin}},\
  and\ \bibinfo {author} {\bibfnamefont {C.}~\bibnamefont {Misbah}},\
  }\bibfield  {title} {\bibinfo {title} {{Three-dimensional vesicles under
  shear flow: Numerical study of dynamics and phase diagram}},\ }\href@noop {}
  {\bibfield  {journal} {\bibinfo  {journal} {Physical Review E}\ ,\ \bibinfo
  {pages} {031921}} (\bibinfo {year} {2011})}\BibitemShut {NoStop}%
\bibitem [{\citenamefont {Zhao}\ and\ \citenamefont
  {Shaqfeh}(2011)}]{zhao-shaqfeh11a}%
  \BibitemOpen
  \bibfield  {author} {\bibinfo {author} {\bibfnamefont {H.}~\bibnamefont
  {Zhao}}\ and\ \bibinfo {author} {\bibfnamefont {E.~S.~G.}\ \bibnamefont
  {Shaqfeh}},\ }\bibfield  {title} {\bibinfo {title} {The dynamics of a vesicle
  in simple shear flow},\ }\href@noop {} {\bibfield  {journal} {\bibinfo
  {journal} {Journal of Fluid Mechanics}\ }\textbf {\bibinfo {volume} {674}},\
  \bibinfo {pages} {578} (\bibinfo {year} {2011})}\BibitemShut {NoStop}%
\bibitem [{\citenamefont {Fedosov}\ \emph {et~al.}(2014)\citenamefont
  {Fedosov}, \citenamefont {Peltomaki},\ and\ \citenamefont
  {Gompper}}]{fedosov-gompper-e14}%
  \BibitemOpen
  \bibfield  {author} {\bibinfo {author} {\bibfnamefont {D.~A.}\ \bibnamefont
  {Fedosov}}, \bibinfo {author} {\bibfnamefont {M.}~\bibnamefont {Peltomaki}},\
  and\ \bibinfo {author} {\bibfnamefont {G.}~\bibnamefont {Gompper}},\
  }\bibfield  {title} {\bibinfo {title} {Deformation and dynamics of red blood
  cells in flow through cylindrical microchannels},\ }\href@noop {} {\bibfield
  {journal} {\bibinfo  {journal} {Soft Matter}\ }\textbf {\bibinfo {volume}
  {10}},\ \bibinfo {pages} {4258} (\bibinfo {year} {2014})}\BibitemShut
  {NoStop}%
\bibitem [{\citenamefont {Farutin}\ and\ \citenamefont
  {Misbah}(2012)}]{farutin-misbah12}%
  \BibitemOpen
  \bibfield  {author} {\bibinfo {author} {\bibfnamefont {A.}~\bibnamefont
  {Farutin}}\ and\ \bibinfo {author} {\bibfnamefont {C.}~\bibnamefont
  {Misbah}},\ }\bibfield  {title} {\bibinfo {title} {Squaring, parity breaking,
  and s tumbling of vesicles under shear flow},\ }\href@noop {} {\bibfield
  {journal} {\bibinfo  {journal} {Physical Review Letters}\ }\textbf {\bibinfo
  {volume} {109}},\ \bibinfo {pages} {248106} (\bibinfo {year}
  {2012})}\BibitemShut {NoStop}%
\bibitem [{\citenamefont {Trozzo}\ \emph {et~al.}(2015)\citenamefont {Trozzo},
  \citenamefont {Boedec}, \citenamefont {Leonetti},\ and\ \citenamefont
  {Jaeger}}]{trozzo-jaeger-e15}%
  \BibitemOpen
  \bibfield  {author} {\bibinfo {author} {\bibfnamefont {R.}~\bibnamefont
  {Trozzo}}, \bibinfo {author} {\bibfnamefont {G.}~\bibnamefont {Boedec}},
  \bibinfo {author} {\bibfnamefont {M.}~\bibnamefont {Leonetti}},\ and\
  \bibinfo {author} {\bibfnamefont {M.}~\bibnamefont {Jaeger}},\ }\bibfield
  {title} {\bibinfo {title} {Axisymmetric boundary element method for vesicles
  in a capillary},\ }\href@noop {} {\bibfield  {journal} {\bibinfo  {journal}
  {J. Comput. Phys.}\ }\textbf {\bibinfo {volume} {289}},\ \bibinfo {pages}
  {62} (\bibinfo {year} {2015})}\BibitemShut {NoStop}%
\bibitem [{\citenamefont {Noireaux}\ and\ \citenamefont
  {Libchaber}(2004)}]{noireaux-libchaber04}%
  \BibitemOpen
  \bibfield  {author} {\bibinfo {author} {\bibfnamefont {V.}~\bibnamefont
  {Noireaux}}\ and\ \bibinfo {author} {\bibfnamefont {A.}~\bibnamefont
  {Libchaber}},\ }\bibfield  {title} {\bibinfo {title} {A vesicle bioreactor as
  a step toward an artificial cell assembly},\ }\href@noop {} {\bibfield
  {journal} {\bibinfo  {journal} {PNAS}\ }\textbf {\bibinfo {volume} {101}},\
  \bibinfo {pages} {17669} (\bibinfo {year} {2004})}\BibitemShut {NoStop}%
\bibitem [{\citenamefont {Karlsson}\ \emph {et~al.}(2004)\citenamefont
  {Karlsson}, \citenamefont {Davidson}, \citenamefont {Karlsson}, \citenamefont
  {Karlsson}, \citenamefont {Bergenholtz}, \citenamefont {Konkoli},
  \citenamefont {Jesorka}, \citenamefont {Lobovkina}, \citenamefont {Hurtig},
  \citenamefont {Voinova},\ and\ \citenamefont {Orwar}}]{karlsson-orwar-e04}%
  \BibitemOpen
  \bibfield  {author} {\bibinfo {author} {\bibfnamefont {M.}~\bibnamefont
  {Karlsson}}, \bibinfo {author} {\bibfnamefont {M.}~\bibnamefont {Davidson}},
  \bibinfo {author} {\bibfnamefont {R.}~\bibnamefont {Karlsson}}, \bibinfo
  {author} {\bibfnamefont {A.}~\bibnamefont {Karlsson}}, \bibinfo {author}
  {\bibfnamefont {J.}~\bibnamefont {Bergenholtz}}, \bibinfo {author}
  {\bibfnamefont {Z.}~\bibnamefont {Konkoli}}, \bibinfo {author} {\bibfnamefont
  {A.}~\bibnamefont {Jesorka}}, \bibinfo {author} {\bibfnamefont
  {T.}~\bibnamefont {Lobovkina}}, \bibinfo {author} {\bibfnamefont
  {J.}~\bibnamefont {Hurtig}}, \bibinfo {author} {\bibfnamefont
  {M.}~\bibnamefont {Voinova}},\ and\ \bibinfo {author} {\bibfnamefont
  {O.}~\bibnamefont {Orwar}},\ }\bibfield  {title} {\bibinfo {title}
  {Biomimetic nanoscale reactors and networks},\ }\href@noop {} {\bibfield
  {journal} {\bibinfo  {journal} {Ann. Rev. Phys. Chem.}\ }\textbf {\bibinfo
  {volume} {55}},\ \bibinfo {pages} {613} (\bibinfo {year} {2004})}\BibitemShut
  {NoStop}%
\bibitem [{\citenamefont {Fendler}(1980)}]{fendler80}%
  \BibitemOpen
  \bibfield  {author} {\bibinfo {author} {\bibfnamefont {J.~H.}\ \bibnamefont
  {Fendler}},\ }\bibfield  {title} {\bibinfo {title} {Surfactant vesicles as
  membrane mimetic agents: Characterization and utilization},\ }\href@noop {}
  {\bibfield  {journal} {\bibinfo  {journal} {Acc. Chem. Res.}\ }\textbf
  {\bibinfo {volume} {13}},\ \bibinfo {pages} {7} (\bibinfo {year}
  {1980})}\BibitemShut {NoStop}%
\bibitem [{\citenamefont {Gregoriadis}(1995)}]{gregoriadis95}%
  \BibitemOpen
  \bibfield  {author} {\bibinfo {author} {\bibfnamefont {G.}~\bibnamefont
  {Gregoriadis}},\ }\bibfield  {title} {\bibinfo {title} {Engineering liposomes
  for drug delivery},\ }\href@noop {} {\bibfield  {journal} {\bibinfo
  {journal} {Trends Biotechnol.}\ }\textbf {\bibinfo {volume} {13}},\ \bibinfo
  {pages} {527} (\bibinfo {year} {1995})}\BibitemShut {NoStop}%
\bibitem [{\citenamefont {Allen}\ and\ \citenamefont
  {Cullis}(2004)}]{allen-cullis04}%
  \BibitemOpen
  \bibfield  {author} {\bibinfo {author} {\bibfnamefont {T.~M.}\ \bibnamefont
  {Allen}}\ and\ \bibinfo {author} {\bibfnamefont {P.~R.}\ \bibnamefont
  {Cullis}},\ }\bibfield  {title} {\bibinfo {title} {Drug delivery systems:
  Entering the mainstream},\ }\href@noop {} {\bibfield  {journal} {\bibinfo
  {journal} {Science}\ }\textbf {\bibinfo {volume} {303}},\ \bibinfo {pages}
  {1818} (\bibinfo {year} {2004})}\BibitemShut {NoStop}%
\bibitem [{\citenamefont {Thomas}\ \emph {et~al.}(2018)\citenamefont {Thomas},
  \citenamefont {O'Brien},\ and\ \citenamefont {Pandit}}]{thomas-pandit-e18}%
  \BibitemOpen
  \bibfield  {author} {\bibinfo {author} {\bibfnamefont {D.}~\bibnamefont
  {Thomas}}, \bibinfo {author} {\bibfnamefont {T.}~\bibnamefont {O'Brien}},\
  and\ \bibinfo {author} {\bibfnamefont {A.}~\bibnamefont {Pandit}},\
  }\bibfield  {title} {\bibinfo {title} {Toward customized extracellular niche
  engineering: Progress in cell-entrapment technologies},\ }\href@noop {}
  {\bibfield  {journal} {\bibinfo  {journal} {Advanced Materials}\ }\textbf
  {\bibinfo {volume} {30}},\ \bibinfo {pages} {1703948} (\bibinfo {year}
  {2018})}\BibitemShut {NoStop}%
\bibitem [{\citenamefont {Skalak}\ and\ \citenamefont
  {Branemark}(2008)}]{skalak-branemark69}%
  \BibitemOpen
  \bibfield  {author} {\bibinfo {author} {\bibfnamefont {R.}~\bibnamefont
  {Skalak}}\ and\ \bibinfo {author} {\bibfnamefont {P.~I.}\ \bibnamefont
  {Branemark}},\ }\bibfield  {title} {\bibinfo {title} {Deformation of red
  blood cells in capillaries},\ }\href@noop {} {\bibfield  {journal} {\bibinfo
  {journal} {Science}\ }\textbf {\bibinfo {volume} {164}},\ \bibinfo {pages}
  {717} (\bibinfo {year} {2008})}\BibitemShut {NoStop}%
\bibitem [{\citenamefont {Chen}\ and\ \citenamefont {Liu}(2012)}]{chan-liu12}%
  \BibitemOpen
  \bibfield  {author} {\bibinfo {author} {\bibfnamefont {C.-K.}\ \bibnamefont
  {Chen}}\ and\ \bibinfo {author} {\bibfnamefont {T.-M.}\ \bibnamefont {Liu}},\
  }\bibfield  {title} {\bibinfo {title} {Imaging morphodynamics of human blood
  cells in vivo with video-rate third harmonic generation microscopy},\
  }\href@noop {} {\bibfield  {journal} {\bibinfo  {journal} {Biomed. Opt.
  Express}\ }\textbf {\bibinfo {volume} {3}},\ \bibinfo {pages} {2860}
  (\bibinfo {year} {2012})}\BibitemShut {NoStop}%
\bibitem [{\citenamefont {Kihm}\ \emph {et~al.}(2018)\citenamefont {Kihm},
  \citenamefont {Kaestner}, \citenamefont {Wagner},\ and\ \citenamefont
  {Quint}}]{kihm-quint-e18}%
  \BibitemOpen
  \bibfield  {author} {\bibinfo {author} {\bibfnamefont {A.}~\bibnamefont
  {Kihm}}, \bibinfo {author} {\bibfnamefont {L.}~\bibnamefont {Kaestner}},
  \bibinfo {author} {\bibfnamefont {C.}~\bibnamefont {Wagner}},\ and\ \bibinfo
  {author} {\bibfnamefont {S.}~\bibnamefont {Quint}},\ }\bibfield  {title}
  {\bibinfo {title} {Classification of red blood cell shapes in flow using
  outlier tolerant machine learning},\ }\href@noop {} {\bibfield  {journal}
  {\bibinfo  {journal} {PLoS Comput. Biol.}\ }\textbf {\bibinfo {volume}
  {14}},\ \bibinfo {pages} {e1006278} (\bibinfo {year} {2018})}\BibitemShut
  {NoStop}%
\bibitem [{\citenamefont {Pivkin}\ \emph {et~al.}(2016)\citenamefont {Pivkin},
  \citenamefont {Peng}, \citenamefont {Karniadakis}, \citenamefont {Buffet},
  \citenamefont {Dao},\ and\ \citenamefont {Suresh}}]{pivkin-suresh-e16}%
  \BibitemOpen
  \bibfield  {author} {\bibinfo {author} {\bibfnamefont {I.}~\bibnamefont
  {Pivkin}}, \bibinfo {author} {\bibfnamefont {Z.}~\bibnamefont {Peng}},
  \bibinfo {author} {\bibfnamefont {G.~E.}\ \bibnamefont {Karniadakis}},
  \bibinfo {author} {\bibfnamefont {P.~A.}\ \bibnamefont {Buffet}}, \bibinfo
  {author} {\bibfnamefont {M.}~\bibnamefont {Dao}},\ and\ \bibinfo {author}
  {\bibfnamefont {S.}~\bibnamefont {Suresh}},\ }\bibfield  {title} {\bibinfo
  {title} {Biomechanics of red blood cells in human spleen and consequences for
  physiology and disease},\ }\href@noop {} {\bibfield  {journal} {\bibinfo
  {journal} {PNAS}\ }\textbf {\bibinfo {volume} {113}},\ \bibinfo {pages}
  {7804} (\bibinfo {year} {2016})}\BibitemShut {NoStop}%
\bibitem [{\citenamefont {Reichel}\ \emph {et~al.}(2019)\citenamefont
  {Reichel}, \citenamefont {Mauer}, \citenamefont {Nawaz}, \citenamefont
  {Gompper}, \citenamefont {Guck},\ and\ \citenamefont
  {Fedosov}}]{reichel-fedosov-e19}%
  \BibitemOpen
  \bibfield  {author} {\bibinfo {author} {\bibfnamefont {F.}~\bibnamefont
  {Reichel}}, \bibinfo {author} {\bibfnamefont {J.}~\bibnamefont {Mauer}},
  \bibinfo {author} {\bibfnamefont {A.~A.}\ \bibnamefont {Nawaz}}, \bibinfo
  {author} {\bibfnamefont {G.}~\bibnamefont {Gompper}}, \bibinfo {author}
  {\bibfnamefont {J.}~\bibnamefont {Guck}},\ and\ \bibinfo {author}
  {\bibfnamefont {D.~A.}\ \bibnamefont {Fedosov}},\ }\bibfield  {title}
  {\bibinfo {title} {High-throughput microfluidic characterization of
  erythrocyte shapes and mechanical variability},\ }\href@noop {} {\bibfield
  {journal} {\bibinfo  {journal} {Biophysical Journal}\ }\textbf {\bibinfo
  {volume} {117}},\ \bibinfo {pages} {14} (\bibinfo {year} {2019})}\BibitemShut
  {NoStop}%
\bibitem [{\citenamefont {Kabacao\u{g}lu}\ and\ \citenamefont
  {Biros}(2019)}]{kabacaoglu-biros19}%
  \BibitemOpen
  \bibfield  {author} {\bibinfo {author} {\bibfnamefont {G.}~\bibnamefont
  {Kabacao\u{g}lu}}\ and\ \bibinfo {author} {\bibfnamefont {G.}~\bibnamefont
  {Biros}},\ }\bibfield  {title} {\bibinfo {title} {Sorting same-size red blood
  cells in deep deterministic lateral displacement devices},\ }\href@noop {}
  {\bibfield  {journal} {\bibinfo  {journal} {Journal of Fluid Mechanics}\
  }\textbf {\bibinfo {volume} {859}},\ \bibinfo {pages} {433} (\bibinfo {year}
  {2019})}\BibitemShut {NoStop}%
\bibitem [{\citenamefont {Zhu}\ \emph {et~al.}(2014)\citenamefont {Zhu},
  \citenamefont {Rorai}, \citenamefont {Mitra},\ and\ \citenamefont
  {Brandt}}]{zhu-brandt-e14}%
  \BibitemOpen
  \bibfield  {author} {\bibinfo {author} {\bibfnamefont {L.}~\bibnamefont
  {Zhu}}, \bibinfo {author} {\bibfnamefont {C.}~\bibnamefont {Rorai}}, \bibinfo
  {author} {\bibfnamefont {D.}~\bibnamefont {Mitra}},\ and\ \bibinfo {author}
  {\bibfnamefont {L.}~\bibnamefont {Brandt}},\ }\bibfield  {title} {\bibinfo
  {title} {A microfluidic device to sort capsules by deformability: a numerical
  study},\ }\href@noop {} {\bibfield  {journal} {\bibinfo  {journal} {Soft
  Matter}\ }\textbf {\bibinfo {volume} {10}},\ \bibinfo {pages} {7705}
  (\bibinfo {year} {2014})}\BibitemShut {NoStop}%
\bibitem [{\citenamefont {Gera}\ \emph {et~al.}(2022)\citenamefont {Gera},
  \citenamefont {Salac},\ and\ \citenamefont {Spagnolie}}]{gera-spagnolie-e22}%
  \BibitemOpen
  \bibfield  {author} {\bibinfo {author} {\bibfnamefont {P.}~\bibnamefont
  {Gera}}, \bibinfo {author} {\bibfnamefont {D.}~\bibnamefont {Salac}},\ and\
  \bibinfo {author} {\bibfnamefont {S.~E.}\ \bibnamefont {Spagnolie}},\
  }\bibfield  {title} {\bibinfo {title} {Swinging and tumbling of
  multicomponent vesicles in flow},\ }\href@noop {} {\bibfield  {journal}
  {\bibinfo  {journal} {Journal of Fluid Mechanics}\ }\textbf {\bibinfo
  {volume} {935}},\ \bibinfo {pages} {A39} (\bibinfo {year}
  {2022})}\BibitemShut {NoStop}%
\bibitem [{\citenamefont {Schmid-Schonbein}\ and\ \citenamefont
  {Wells}(1969)}]{schonbein-wells69}%
  \BibitemOpen
  \bibfield  {author} {\bibinfo {author} {\bibfnamefont {H.}~\bibnamefont
  {Schmid-Schonbein}}\ and\ \bibinfo {author} {\bibfnamefont {R.}~\bibnamefont
  {Wells}},\ }\bibfield  {title} {\bibinfo {title} {Fluid drop-like transition
  of erythrocytes under shear},\ }\href@noop {} {\bibfield  {journal} {\bibinfo
   {journal} {Science}\ }\textbf {\bibinfo {volume} {165}},\ \bibinfo {pages}
  {288} (\bibinfo {year} {1969})}\BibitemShut {NoStop}%
\bibitem [{\citenamefont {Goldsmith}\ \emph {et~al.}(1972)\citenamefont
  {Goldsmith}, \citenamefont {Marlow},\ and\ \citenamefont
  {MacIntosh}}]{goldsmith-macintosh-e72}%
  \BibitemOpen
  \bibfield  {author} {\bibinfo {author} {\bibfnamefont {H.~L.}\ \bibnamefont
  {Goldsmith}}, \bibinfo {author} {\bibfnamefont {J.}~\bibnamefont {Marlow}},\
  and\ \bibinfo {author} {\bibfnamefont {F.~C.}\ \bibnamefont {MacIntosh}},\
  }\bibfield  {title} {\bibinfo {title} {Flow behaviour of erythrocytes - i.
  rotation and deformation in dilute suspensions},\ }\href@noop {} {\bibfield
  {journal} {\bibinfo  {journal} {Proc. R. Soc. London B}\ }\textbf {\bibinfo
  {volume} {182}},\ \bibinfo {pages} {351} (\bibinfo {year}
  {1972})}\BibitemShut {NoStop}%
\bibitem [{\citenamefont {Zhao}\ and\ \citenamefont
  {Shaqfeh}(2009)}]{zhao-shaqfeh09}%
  \BibitemOpen
  \bibfield  {author} {\bibinfo {author} {\bibfnamefont {H.}~\bibnamefont
  {Zhao}}\ and\ \bibinfo {author} {\bibfnamefont {E.~S.~G.}\ \bibnamefont
  {Shaqfeh}},\ }\href@noop {} {\emph {\bibinfo {title} {The dynamics of a
  vesicle in shear flow}}},\ \bibinfo {type} {Tech. Rep.}\ (\bibinfo
  {institution} {Stanford University},\ \bibinfo {year} {2009})\BibitemShut
  {NoStop}%
\bibitem [{\citenamefont {Noguchi}\ and\ \citenamefont
  {Gompper}(2007)}]{noguchi-gompper07}%
  \BibitemOpen
  \bibfield  {author} {\bibinfo {author} {\bibfnamefont {H.}~\bibnamefont
  {Noguchi}}\ and\ \bibinfo {author} {\bibfnamefont {D.~G.}\ \bibnamefont
  {Gompper}},\ }\bibfield  {title} {\bibinfo {title} {Swinging and tumbling of
  fluid vesicles in shear flow},\ }\href@noop {} {\bibfield  {journal}
  {\bibinfo  {journal} {Physical Review Letters}\ }\textbf {\bibinfo {volume}
  {98}},\ \bibinfo {pages} {128103} (\bibinfo {year} {2007})}\BibitemShut
  {NoStop}%
\bibitem [{\citenamefont {Kaoui}\ \emph
  {et~al.}(2009{\natexlab{b}})\citenamefont {Kaoui}, \citenamefont {Farutin},\
  and\ \citenamefont {Misbah}}]{kaoui-misbah-e09b}%
  \BibitemOpen
  \bibfield  {author} {\bibinfo {author} {\bibfnamefont {B.}~\bibnamefont
  {Kaoui}}, \bibinfo {author} {\bibfnamefont {A.}~\bibnamefont {Farutin}},\
  and\ \bibinfo {author} {\bibfnamefont {C.}~\bibnamefont {Misbah}},\
  }\bibfield  {title} {\bibinfo {title} {Vesicles under simple shear flow:
  Elucidating the role of relevant control parameters},\ }\href@noop {}
  {\bibfield  {journal} {\bibinfo  {journal} {Physical Review E}\ }\textbf
  {\bibinfo {volume} {80}},\ \bibinfo {pages} {061905} (\bibinfo {year}
  {2009}{\natexlab{b}})}\BibitemShut {NoStop}%
\bibitem [{\citenamefont {Danker}\ \emph {et~al.}(2009)\citenamefont {Danker},
  \citenamefont {Vlahovska},\ and\ \citenamefont {Misbah}}]{danker-misbah-e09}%
  \BibitemOpen
  \bibfield  {author} {\bibinfo {author} {\bibfnamefont {G.}~\bibnamefont
  {Danker}}, \bibinfo {author} {\bibfnamefont {P.~M.}\ \bibnamefont
  {Vlahovska}},\ and\ \bibinfo {author} {\bibfnamefont {C.}~\bibnamefont
  {Misbah}},\ }\bibfield  {title} {\bibinfo {title} {Vesicle in poiseuille
  flow},\ }\href@noop {} {\bibfield  {journal} {\bibinfo  {journal} {Physical
  Review Letters}\ }\textbf {\bibinfo {volume} {102}},\ \bibinfo {pages}
  {148102} (\bibinfo {year} {2009})}\BibitemShut {NoStop}%
\bibitem [{\citenamefont {Deschamps}\ \emph {et~al.}(2009)\citenamefont
  {Deschamps}, \citenamefont {Kantsler},\ and\ \citenamefont
  {Steinberg}}]{deschamps-steinberg-e09}%
  \BibitemOpen
  \bibfield  {author} {\bibinfo {author} {\bibfnamefont {J.}~\bibnamefont
  {Deschamps}}, \bibinfo {author} {\bibfnamefont {V.}~\bibnamefont
  {Kantsler}},\ and\ \bibinfo {author} {\bibfnamefont {V.}~\bibnamefont
  {Steinberg}},\ }\bibfield  {title} {\bibinfo {title} {Phase diagram of single
  vesicle dynamical states in shear flow},\ }\href@noop {} {\bibfield
  {journal} {\bibinfo  {journal} {Phys. Rev. Lett.}\ }\textbf {\bibinfo
  {volume} {102}},\ \bibinfo {pages} {118105} (\bibinfo {year}
  {2009})}\BibitemShut {NoStop}%
\bibitem [{\citenamefont {Kaoui}\ \emph {et~al.}(2012)\citenamefont {Kaoui},
  \citenamefont {Kr\"{u}ger},\ and\ \citenamefont
  {Harting}}]{kaoui-harting-e12}%
  \BibitemOpen
  \bibfield  {author} {\bibinfo {author} {\bibfnamefont {B.}~\bibnamefont
  {Kaoui}}, \bibinfo {author} {\bibfnamefont {T.}~\bibnamefont {Kr\"{u}ger}},\
  and\ \bibinfo {author} {\bibfnamefont {J.}~\bibnamefont {Harting}},\
  }\bibfield  {title} {\bibinfo {title} {How does confinement affect the
  dynamics of viscous vesicles and red blood cells?},\ }\href@noop {}
  {\bibfield  {journal} {\bibinfo  {journal} {Soft Matter}\ }\textbf {\bibinfo
  {volume} {8}},\ \bibinfo {pages} {9246} (\bibinfo {year} {2012})}\BibitemShut
  {NoStop}%
\bibitem [{\citenamefont {Quint}\ \emph {et~al.}(2017)\citenamefont {Quint},
  \citenamefont {Christ}, \citenamefont {Guckenberger}, \citenamefont
  {Himbert}, \citenamefont {Kaestner}, \citenamefont {Gekle},\ and\
  \citenamefont {Wagner}}]{quint-wagner-e17}%
  \BibitemOpen
  \bibfield  {author} {\bibinfo {author} {\bibfnamefont {S.}~\bibnamefont
  {Quint}}, \bibinfo {author} {\bibfnamefont {A.~F.}\ \bibnamefont {Christ}},
  \bibinfo {author} {\bibfnamefont {A.}~\bibnamefont {Guckenberger}}, \bibinfo
  {author} {\bibfnamefont {S.}~\bibnamefont {Himbert}}, \bibinfo {author}
  {\bibfnamefont {L.}~\bibnamefont {Kaestner}}, \bibinfo {author}
  {\bibfnamefont {S.}~\bibnamefont {Gekle}},\ and\ \bibinfo {author}
  {\bibfnamefont {C.}~\bibnamefont {Wagner}},\ }\bibfield  {title} {\bibinfo
  {title} {3d tomogrophy of cells in micro-channels},\ }\href@noop {}
  {\bibfield  {journal} {\bibinfo  {journal} {Appl. Phys. Lett.}\ }\textbf
  {\bibinfo {volume} {111}},\ \bibinfo {pages} {103701} (\bibinfo {year}
  {2017})}\BibitemShut {NoStop}%
\bibitem [{\citenamefont {Guckenberger}\ \emph {et~al.}(2018)\citenamefont
  {Guckenberger}, \citenamefont {Kihm}, \citenamefont {John}, \citenamefont
  {Wagner},\ and\ \citenamefont {Gekle}}]{guckenberger-gekle-e18}%
  \BibitemOpen
  \bibfield  {author} {\bibinfo {author} {\bibfnamefont {A.}~\bibnamefont
  {Guckenberger}}, \bibinfo {author} {\bibfnamefont {A.}~\bibnamefont {Kihm}},
  \bibinfo {author} {\bibfnamefont {T.}~\bibnamefont {John}}, \bibinfo {author}
  {\bibfnamefont {C.}~\bibnamefont {Wagner}},\ and\ \bibinfo {author}
  {\bibfnamefont {S.}~\bibnamefont {Gekle}},\ }\bibfield  {title} {\bibinfo
  {title} {Numerical-experimental observation of shape bistability of red blood
  cells flowing in a microchannel},\ }\href@noop {} {\bibfield  {journal}
  {\bibinfo  {journal} {Soft Matter}\ }\textbf {\bibinfo {volume} {14}},\
  \bibinfo {pages} {2032} (\bibinfo {year} {2018})}\BibitemShut {NoStop}%
\bibitem [{\citenamefont {Farutin}\ \emph {et~al.}(2014)\citenamefont
  {Farutin}, \citenamefont {Biben},\ and\ \citenamefont
  {Misbah}}]{farutin-misbah-e14}%
  \BibitemOpen
  \bibfield  {author} {\bibinfo {author} {\bibfnamefont {A.}~\bibnamefont
  {Farutin}}, \bibinfo {author} {\bibfnamefont {T.}~\bibnamefont {Biben}},\
  and\ \bibinfo {author} {\bibfnamefont {C.}~\bibnamefont {Misbah}},\
  }\bibfield  {title} {\bibinfo {title} {3d numerical simulations of vesicle
  and inextensible capsule dynamics},\ }\href@noop {} {\bibfield  {journal}
  {\bibinfo  {journal} {J. Comput. Phys}\ }\textbf {\bibinfo {volume} {275}},\
  \bibinfo {pages} {539} (\bibinfo {year} {2014})}\BibitemShut {NoStop}%
\bibitem [{\citenamefont {Farutin}\ \emph {et~al.}(2016)\citenamefont
  {Farutin}, \citenamefont {Piasecki}, \citenamefont {Slowicka}, \citenamefont
  {Misbah}, \citenamefont {Wajnryb},\ and\ \citenamefont
  {Ekiel-Jezewska}}]{farutin-jezewska-e16}%
  \BibitemOpen
  \bibfield  {author} {\bibinfo {author} {\bibfnamefont {A.}~\bibnamefont
  {Farutin}}, \bibinfo {author} {\bibfnamefont {T.}~\bibnamefont {Piasecki}},
  \bibinfo {author} {\bibfnamefont {A.~M.}\ \bibnamefont {Slowicka}}, \bibinfo
  {author} {\bibfnamefont {C.}~\bibnamefont {Misbah}}, \bibinfo {author}
  {\bibfnamefont {E.}~\bibnamefont {Wajnryb}},\ and\ \bibinfo {author}
  {\bibfnamefont {M.~L.}\ \bibnamefont {Ekiel-Jezewska}},\ }\bibfield  {title}
  {\bibinfo {title} {Dynamics of flexible fibers and vesicles in poiseuille
  flow at low reynolds number},\ }\href@noop {} {\bibfield  {journal} {\bibinfo
   {journal} {Soft Matter}\ }\textbf {\bibinfo {volume} {12}},\ \bibinfo
  {pages} {7307} (\bibinfo {year} {2016})}\BibitemShut {NoStop}%
\bibitem [{\citenamefont {Ebrahimi}\ \emph {et~al.}(2021)\citenamefont
  {Ebrahimi}, \citenamefont {Balogh},\ and\ \citenamefont
  {Bagchi}}]{ebrahimi-bagchi-e21}%
  \BibitemOpen
  \bibfield  {author} {\bibinfo {author} {\bibfnamefont {E.}~\bibnamefont
  {Ebrahimi}}, \bibinfo {author} {\bibfnamefont {P.}~\bibnamefont {Balogh}},\
  and\ \bibinfo {author} {\bibfnamefont {P.}~\bibnamefont {Bagchi}},\
  }\bibfield  {title} {\bibinfo {title} {{Motion of a capsule in a curved
  tube}},\ }\href@noop {} {\bibfield  {journal} {\bibinfo  {journal} {Journal
  of Fluid Mechanics}\ }\textbf {\bibinfo {volume} {907}},\ \bibinfo {pages}
  {A28} (\bibinfo {year} {2021})}\BibitemShut {NoStop}%
\bibitem [{\citenamefont {Cordasco}\ \emph {et~al.}(2014)\citenamefont
  {Cordasco}, \citenamefont {Yazdani},\ and\ \citenamefont
  {Bagchi}}]{cordasco-bagchi-e14}%
  \BibitemOpen
  \bibfield  {author} {\bibinfo {author} {\bibfnamefont {D.}~\bibnamefont
  {Cordasco}}, \bibinfo {author} {\bibfnamefont {A.}~\bibnamefont {Yazdani}},\
  and\ \bibinfo {author} {\bibfnamefont {P.}~\bibnamefont {Bagchi}},\
  }\bibfield  {title} {\bibinfo {title} {Comparison of erythrocyte dynamics in
  shear flow under different stress-free configurations},\ }\href@noop {}
  {\bibfield  {journal} {\bibinfo  {journal} {Physics of Fluids}\ }\textbf
  {\bibinfo {volume} {26}},\ \bibinfo {pages} {041902} (\bibinfo {year}
  {2014})}\BibitemShut {NoStop}%
\bibitem [{\citenamefont {Ghigliotti}\ \emph {et~al.}(2011)\citenamefont
  {Ghigliotti}, \citenamefont {Rahimian}, \citenamefont {Biros},\ and\
  \citenamefont {Misbah}}]{ghigliotti-misbah-e10}%
  \BibitemOpen
  \bibfield  {author} {\bibinfo {author} {\bibfnamefont {G.}~\bibnamefont
  {Ghigliotti}}, \bibinfo {author} {\bibfnamefont {A.}~\bibnamefont
  {Rahimian}}, \bibinfo {author} {\bibfnamefont {G.}~\bibnamefont {Biros}},\
  and\ \bibinfo {author} {\bibfnamefont {C.}~\bibnamefont {Misbah}},\
  }\bibfield  {title} {\bibinfo {title} {Vesicle migration and spatial
  organization driven by flow line curvature},\ }\href@noop {} {\bibfield
  {journal} {\bibinfo  {journal} {Physical Review Letters}\ }\textbf {\bibinfo
  {volume} {106}},\ \bibinfo {pages} {028101} (\bibinfo {year}
  {2011})}\BibitemShut {NoStop}%
\bibitem [{\citenamefont {Vlahovska}\ \emph {et~al.}(2009)\citenamefont
  {Vlahovska}, \citenamefont {Podgorski},\ and\ \citenamefont
  {Misbah}}]{vlahovska-misbah-e09}%
  \BibitemOpen
  \bibfield  {author} {\bibinfo {author} {\bibfnamefont {P.~M.}\ \bibnamefont
  {Vlahovska}}, \bibinfo {author} {\bibfnamefont {T.}~\bibnamefont
  {Podgorski}},\ and\ \bibinfo {author} {\bibfnamefont {C.}~\bibnamefont
  {Misbah}},\ }\bibfield  {title} {\bibinfo {title} {Vesicles and red blood
  cells in flow: From individual dynamics to rheology},\ }\href@noop {}
  {\bibfield  {journal} {\bibinfo  {journal} {C. R. Physique}\ }\textbf
  {\bibinfo {volume} {10}},\ \bibinfo {pages} {775} (\bibinfo {year}
  {2009})}\BibitemShut {NoStop}%
\bibitem [{\citenamefont {Kantsler}\ and\ \citenamefont
  {Steinberg}(2005)}]{kantsler-steinberg-e05}%
  \BibitemOpen
  \bibfield  {author} {\bibinfo {author} {\bibfnamefont {V.}~\bibnamefont
  {Kantsler}}\ and\ \bibinfo {author} {\bibfnamefont {V.}~\bibnamefont
  {Steinberg}},\ }\bibfield  {title} {\bibinfo {title} {Orientation and
  dynamics of a vesicle in tank-treading motion in shear flow},\ }\href@noop {}
  {\bibfield  {journal} {\bibinfo  {journal} {Physical Review Letters}\
  }\textbf {\bibinfo {volume} {95}},\ \bibinfo {pages} {258101} (\bibinfo
  {year} {2005})}\BibitemShut {NoStop}%
\bibitem [{\citenamefont {Kantsler}\ and\ \citenamefont
  {Steinberg}(2006)}]{kantsler-steinberg-e06}%
  \BibitemOpen
  \bibfield  {author} {\bibinfo {author} {\bibfnamefont {V.}~\bibnamefont
  {Kantsler}}\ and\ \bibinfo {author} {\bibfnamefont {V.}~\bibnamefont
  {Steinberg}},\ }\bibfield  {title} {\bibinfo {title} {Transition to tumbling
  and two regimes of tumbling motion of a vesicle in shear flow},\ }\href@noop
  {} {\bibfield  {journal} {\bibinfo  {journal} {Physical Review Letters}\
  }\textbf {\bibinfo {volume} {96}} (\bibinfo {year} {2006})}\BibitemShut
  {NoStop}%
\bibitem [{\citenamefont {Nait-Ouhra}\ \emph {et~al.}(2018)\citenamefont
  {Nait-Ouhra}, \citenamefont {Guckenberger}, \citenamefont {Farutin},
  \citenamefont {Ez-Zahraouy}, \citenamefont {Benyoussef}, \citenamefont
  {Gekle},\ and\ \citenamefont {Misbah}}]{ouhra-misbah-e18}%
  \BibitemOpen
  \bibfield  {author} {\bibinfo {author} {\bibfnamefont {A.}~\bibnamefont
  {Nait-Ouhra}}, \bibinfo {author} {\bibfnamefont {A.}~\bibnamefont
  {Guckenberger}}, \bibinfo {author} {\bibfnamefont {A.}~\bibnamefont
  {Farutin}}, \bibinfo {author} {\bibfnamefont {H.}~\bibnamefont
  {Ez-Zahraouy}}, \bibinfo {author} {\bibfnamefont {A.}~\bibnamefont
  {Benyoussef}}, \bibinfo {author} {\bibfnamefont {S.}~\bibnamefont {Gekle}},\
  and\ \bibinfo {author} {\bibfnamefont {C.}~\bibnamefont {Misbah}},\
  }\bibfield  {title} {\bibinfo {title} {Lateral vesicle migration in a bounded
  shear flow: Viscosity contrast leads to off-centered solutions},\ }\href@noop
  {} {\bibfield  {journal} {\bibinfo  {journal} {Physical Review Fluids}\
  }\textbf {\bibinfo {volume} {3}},\ \bibinfo {pages} {123601} (\bibinfo {year}
  {2018})}\BibitemShut {NoStop}%
\bibitem [{\citenamefont {S.}(1923)}]{taylor1923}%
  \BibitemOpen
  \bibfield  {author} {\bibinfo {author} {\bibfnamefont {G.~I. T. F.~R.}\
  \bibnamefont {S.}},\ }\bibfield  {title} {\bibinfo {title} {{LXXV. On the
  decay of vortices in a viscous fluid}},\ }\href@noop {} {\bibfield  {journal}
  {\bibinfo  {journal} {The London, Edinburgh, and Dublin Philosophical
  Magazine and Journal of Science}\ }\textbf {\bibinfo {volume} {46}},\
  \bibinfo {pages} {671} (\bibinfo {year} {1923})}\BibitemShut {NoStop}%
\bibitem [{\citenamefont {Kaoui}\ \emph {et~al.}(2008)\citenamefont {Kaoui},
  \citenamefont {Ristow}, \citenamefont {Cantat}, \citenamefont {Misbah},\ and\
  \citenamefont {Zimmermann}}]{kaoui-zimmermann-08}%
  \BibitemOpen
  \bibfield  {author} {\bibinfo {author} {\bibfnamefont {B.}~\bibnamefont
  {Kaoui}}, \bibinfo {author} {\bibfnamefont {G.~H.}\ \bibnamefont {Ristow}},
  \bibinfo {author} {\bibfnamefont {I.}~\bibnamefont {Cantat}}, \bibinfo
  {author} {\bibfnamefont {C.}~\bibnamefont {Misbah}},\ and\ \bibinfo {author}
  {\bibfnamefont {W.}~\bibnamefont {Zimmermann}},\ }\bibfield  {title}
  {\bibinfo {title} {{Lateral migration of a two-dimensional vesicle in
  unbounded Poiseuille flow}},\ }\href@noop {} {\bibfield  {journal} {\bibinfo
  {journal} {{Physical Review E}}\ }\textbf {\bibinfo {volume} {{77}}}
  (\bibinfo {year} {{2008}})}\BibitemShut {NoStop}%
\bibitem [{\citenamefont {Kabacao\u{g}lu}\ \emph {et~al.}(2018)\citenamefont
  {Kabacao\u{g}lu}, \citenamefont {Quaife},\ and\ \citenamefont
  {Biros}}]{kabacaoglu-biros-e18}%
  \BibitemOpen
  \bibfield  {author} {\bibinfo {author} {\bibfnamefont {G.}~\bibnamefont
  {Kabacao\u{g}lu}}, \bibinfo {author} {\bibfnamefont {B.}~\bibnamefont
  {Quaife}},\ and\ \bibinfo {author} {\bibfnamefont {G.}~\bibnamefont
  {Biros}},\ }\bibfield  {title} {\bibinfo {title} {Low-resolution simulations
  of vesicle suspensions in 2{D}},\ }\href@noop {} {\bibfield  {journal}
  {\bibinfo  {journal} {Journal of Computational Physics}\ }\textbf {\bibinfo
  {volume} {357}},\ \bibinfo {pages} {43} (\bibinfo {year} {2018})}\BibitemShut
  {NoStop}%
\bibitem [{\citenamefont {Fung}(1990)}]{fung90}%
  \BibitemOpen
  \bibfield  {author} {\bibinfo {author} {\bibfnamefont {Y.}~\bibnamefont
  {Fung}},\ }\href@noop {} {\emph {\bibinfo {title} {Biomechanics}}}\ (\bibinfo
   {publisher} {Springer},\ \bibinfo {address} {New York},\ \bibinfo {year}
  {1990})\BibitemShut {NoStop}%
\bibitem [{\citenamefont {Mohandas}\ and\ \citenamefont
  {Evans}(1994)}]{mohandas-evans94}%
  \BibitemOpen
  \bibfield  {author} {\bibinfo {author} {\bibfnamefont {N.}~\bibnamefont
  {Mohandas}}\ and\ \bibinfo {author} {\bibfnamefont {E.}~\bibnamefont
  {Evans}},\ }\bibfield  {title} {\bibinfo {title} {Mechanical properties of
  the red blood cell membrane in relation to molecular structure and genetic
  defects},\ }\href@noop {} {\bibfield  {journal} {\bibinfo  {journal} {Annu.
  Rev. Biophys. Biomol. Struct.}\ }\textbf {\bibinfo {volume} {23}},\ \bibinfo
  {pages} {787} (\bibinfo {year} {1994})}\BibitemShut {NoStop}%
\bibitem [{\citenamefont {Kaoui}\ \emph {et~al.}(2011)\citenamefont {Kaoui},
  \citenamefont {Tahiri}, \citenamefont {Biben}, \citenamefont {Ez-Zahraouy},
  \citenamefont {Benyoussef}, \citenamefont {Biros},\ and\ \citenamefont
  {Misbah}}]{kaoui-misbah-e11}%
  \BibitemOpen
  \bibfield  {author} {\bibinfo {author} {\bibfnamefont {B.}~\bibnamefont
  {Kaoui}}, \bibinfo {author} {\bibfnamefont {N.}~\bibnamefont {Tahiri}},
  \bibinfo {author} {\bibfnamefont {T.}~\bibnamefont {Biben}}, \bibinfo
  {author} {\bibfnamefont {H.}~\bibnamefont {Ez-Zahraouy}}, \bibinfo {author}
  {\bibfnamefont {A.}~\bibnamefont {Benyoussef}}, \bibinfo {author}
  {\bibfnamefont {G.}~\bibnamefont {Biros}},\ and\ \bibinfo {author}
  {\bibfnamefont {C.}~\bibnamefont {Misbah}},\ }\bibfield  {title} {\bibinfo
  {title} {{Complexity of vesicle microcirculation}},\ }\href@noop {}
  {\bibfield  {journal} {\bibinfo  {journal} {Physical Review E}\ ,\ \bibinfo
  {pages} {041906}} (\bibinfo {year} {2011})}\BibitemShut {NoStop}%
\bibitem [{\citenamefont {Seifert}(1991)}]{seifert91}%
  \BibitemOpen
  \bibfield  {author} {\bibinfo {author} {\bibfnamefont {U.}~\bibnamefont
  {Seifert}},\ }\bibfield  {title} {\bibinfo {title} {Adhesion of vesicles in
  two-dimensions},\ }\href@noop {} {\bibfield  {journal} {\bibinfo  {journal}
  {Physical Review A}\ }\textbf {\bibinfo {volume} {43}},\ \bibinfo {pages}
  {6803} (\bibinfo {year} {1991})}\BibitemShut {NoStop}%
\bibitem [{\citenamefont {Agarwal}\ and\ \citenamefont
  {Biros}(2020)}]{agarwal-biros20}%
  \BibitemOpen
  \bibfield  {author} {\bibinfo {author} {\bibfnamefont {D.}~\bibnamefont
  {Agarwal}}\ and\ \bibinfo {author} {\bibfnamefont {G.}~\bibnamefont
  {Biros}},\ }\bibfield  {title} {\bibinfo {title} {{Stable shapes of
  three-dimensional vesicles in unconfined and confined Poiseuille flow}},\
  }\href@noop {} {\bibfield  {journal} {\bibinfo  {journal} {{Physical Review
  Fluids}}\ }\textbf {\bibinfo {volume} {{5}}},\ \bibinfo {pages} {{013603}}
  (\bibinfo {year} {{2020}})}\BibitemShut {NoStop}%
\bibitem [{\citenamefont {Rahimian}\ \emph {et~al.}(2010)\citenamefont
  {Rahimian}, \citenamefont {Veerapaneni},\ and\ \citenamefont
  {Biros}}]{rahimian-biros-e10}%
  \BibitemOpen
  \bibfield  {author} {\bibinfo {author} {\bibfnamefont {A.}~\bibnamefont
  {Rahimian}}, \bibinfo {author} {\bibfnamefont {S.~K.}\ \bibnamefont
  {Veerapaneni}},\ and\ \bibinfo {author} {\bibfnamefont {G.}~\bibnamefont
  {Biros}},\ }\bibfield  {title} {\bibinfo {title} {Dynamic simulation of
  locally inextensible vesicles suspended in an arbitrary two-dimensional
  domain, a boundary integral method},\ }\href@noop {} {\bibfield  {journal}
  {\bibinfo  {journal} {Journal of Computational Physics}\ }\textbf {\bibinfo
  {volume} {229}},\ \bibinfo {pages} {6466} (\bibinfo {year}
  {2010})}\BibitemShut {NoStop}%
\end{thebibliography}%

\end{document}